\documentclass[iop]{emulateapj}
\usepackage{amsmath}

\newcommand{\kms}{\ifmmode {\rm km\ s}^{-1} \else km s$^{-1}$\fi}
\newcommand{\Msun}{\ifmmode {\rm M}_{\odot} \else M$_{\odot}$\fi}
\newcommand{\Lsun}{\ifmmode {\rm L}_{\odot} \else L$_{\odot}$\fi}
\newcommand{\qo}{\ifmmode q_{\rm o} \else $q_{\rm o}$\fi}
\newcommand{\Ho}{\ifmmode H_{\rm o} \else $H_{\rm o}$\fi}
\newcommand{\ho}{\ifmmode h_{\rm o} \else $h_{\rm o}$\fi}

\newcommand{\vFWHM}{\ifmmode v_{\mbox{\tiny FWHM}} \else
                    $v_{\mbox{\tiny FWHM}}$\fi}
\newcommand{\CCF}{\ifmmode F_{\it CCF} \else $F_{\it CCF}$\fi}
\newcommand{\ACF}{\ifmmode F_{\it ACF} \else $F_{\it ACF}$\fi}
\newcommand{\Halpha}{\ifmmode {\rm H}\alpha \else H$\alpha$\fi}
\newcommand{\Hbeta}{\ifmmode {\rm H}\beta \else H$\beta$\fi}
\newcommand{\Hgamma}{\ifmmode {\rm H}\gamma \else H$\gamma$\fi}
\newcommand{\Hdelta}{\ifmmode {\rm H}\delta \else H$\delta$\fi}
\newcommand{\Lya}{\ifmmode {\rm Ly}\alpha \else Ly$\alpha$\fi}
\newcommand{\Lyb}{\ifmmode {\rm Ly}\beta \else Ly$\beta$\fi}
\newcommand{\HeI}{\ifmmode {\rm He}\,{\sc i}\,\lambda5876 \else 
	          He\,{\sc i}\,$\lambda5876$\fi}
\newcommand{\HeII}{\ifmmode {\rm He}\,{\sc ii}\,\lambda4686 \else 
	           He\,{\sc ii}\,$\lambda4686$\fi}

\newcommand{\heii}{He\,{\sc ii}}

\newcommand{\ciii}{\ifmmode {\rm C}\,{\sc iii} \else C\,{\sc iii}\fi}
\newcommand{\civ}{\ifmmode {\rm C}\,{\sc iv} \else C\,{\sc iv}\fi}
\newcommand{\CIV}{\ifmmode {\rm C}\,{\sc iv}\,\lambda1549 \else 
	           C\,{\sc iv}\,$\lambda1549$\fi}

\newcommand{\oiii}{O\,{\sc iii}}
\newcommand{\ob}{[O\,{\sc iii}]\,$\lambda \lambda 4959,5007$}

\newcommand{\mgii}{Mg\,{\sc ii}}

\newcommand{\civm}{{\rm C}\,{\scriptstyle{\rm IV}}}
\newcommand{\aam}{\mbox{\normalfont\AA}}

\shorttitle{CIV Line Shape}
\shortauthors{}

\received{31 July 2012}
\accepted{17 Sept 2012}

\begin{document}

\title{Are Outflows Biasing Single-Epoch C\,{\small IV} Black Hole Mass Estimates?}

\author{ K.~D.~Denney\altaffilmark{1,2}}

\altaffiltext{1}{Dark Cosmology Centre, 
                 Niels Bohr Institute, 
                 Juliane Maries Vej 30, 2100 Copenhagen \O, Denmark;
                 kelly@dark-cosmology.dk}


\footnotetext[2]{Marie Curie Fellow}

\begin{abstract}

   We use a combination of reverberation mapping data and single-epoch
  spectra of the \civ\ emission line in a sample of both low and
  high-redshift active galactic nuclei (AGNs) to investigate sources of
  the discrepancies between \civ- and \Hbeta-based single-epoch black
  hole mass estimates.  We find that for all reverberation mapped
  sources, there is a component of the line profile that does not
  reverberate, and the velocity characteristics of this component vary
  from object-to-object.  The differing strength and properties of this
  non-variable component are responsible for much of the scatter in
  \civ-based black hole masses compared to \Hbeta\ masses.  The \civ\
  mass bias introduced by this non-variable component is correlated with
  the shape of the \civ\ line, allowing us to make an empirical
  correction to the black hole mass estimates.  Using this correction
  and accounting for other sources of scatter such as poor data quality
  and data inhomogeneity reduces the scatter between the \civ\ and
  \Hbeta\ masses in our sample by a factor of $\sim$2, to only $\sim$0.2
  dex.  We discuss the possibility that this non-variable \civ\
  component originates in an orientation-dependent outflow from either
  the proposed broad line region (BLR) disk-wind or the intermediate
  line region (ILR), a high-velocity inner extension of the narrow line
  region (NLR).
 
\end{abstract}



\section{INTRODUCTION}
\label{S_Intro}

Observations of galaxies in the local universe have led to the nearly
universal acceptance that massive galaxies house a supermassive black
hole (BH) at the center of their gravitational potential well.  In the
local universe, the masses of the BHs are strongly correlated with some
properties of their host galaxies, leading to the $M_{\rm BH}-\sigma_*$
and $M_{\rm BH}-L_{\rm bulge}$ relationships \citep[e.g.][]{Ferrarese00,
  Gebhardt00a, Tremaine02, Marconi03, Graham07, Gultekin09, Graham11}.
Because of the apparent tight evolutionary relationship between the BH
and its host, one way in which to probe galaxy evolution is to trace the
growth of the central BH.  In order to do this, we need reliable methods
for measuring BH masses in the near {\it and} distant universe.

Direct dynamical BH mass measurement methods work only for nearby
quiescent galaxies in which the gravitational sphere of influence of the
BH can be spatially resolved (see, e.g., the review by
\citealp{Ferrarese05}, though also see \citealp{Gultekin09} for a
discussion of deriving masses without resolving the sphere of
influence).  On the other hand, reverberation mapping
\citep{Blandford82, Peterson93} can be applied to active galactic nuclei
(AGNs) to measure BH masses both locally and at cosmological distances.
Unfortunately, the required length of the spectrophotometric monitoring
campaigns needed to make the measurements becomes longer with increasing
source luminosity and distance.  As a result, current measurements only
extend to sources with $z \lesssim 0.3$ (see, e.g., \citealp{Peterson04,
  Bentz09lamp, Denney10, Grier12b}; see Kaspi et al.\ 2007 for a
tentative result for a $z\sim$2 quasar).

Extending direct mass measurements to large samples of objects at high
redshifts in order to study the co-evolution of BHs and galaxies is not
likely to be possible in the near future with current resources.
However, the results from reverberation mapping of local AGNs can be
used to calibrate {\it indirect} estimates of BH masses for large
samples of objects using single spectra.  These ``single-epoch'' (SE)
mass estimates rely on the observed tight correlation between the
monochromatic AGN luminosity, $L$, and the radius of the broad line
region (BLR), $R$, measured in reverberation mapping experiments.  With
this method, only two observables are then needed: (1) a measurement of
the AGN luminosity to use as a proxy for the BLR radius through this
$R-L$ relationship \citep{Kaspi05, Kaspi07, Bentz06a, Bentz09rl}, and (2)
a measurement of a broad emission-line width (typically H$\alpha$,
H$\beta$, \mgii\,$\lambda1549$, or \CIV) to estimate the
Doppler-broadened BLR line-of-sight (LOS) velocity dispersion, $V$.
Combining these observables, a virial BH mass can be calculated as
$M_{\rm BH}=fRV^2/G$, where $G$ is the gravitational constant and $f$ is
a scale factor of order unity related to the BLR geometry and
kinematics\footnote{For reverberation mapping results that use the line
  width measured from the rms spectrum, an ensemble average, $\langle f
  \rangle$, sets the AGN BH mass scale zero-point, and is determined
  assuming AGNs follow the same $M-\sigma_*$ relationship as quiescent
  galaxies \citep{Onken04, Woo10}.  For SE masses, $f$ has a different
  calibration based on the differences between the SE and rms line
  widths for H$\beta$ \citep[see][]{Collin06}.}.

The majority of reverberation mapping results to date use time delays
measured from \Hbeta\ emission line variability.  Consequently, the most
robust SE mass scaling relation is based directly on the $R-L$
relationship for this emission-line \citep[see][]{Collin06,
  Vestergaard06, Bentz09rl}.  Scaling relationships for other broad
emission lines are also available \citep{McLure02, Vestergaard06,
  McGill08, Vestergaard&Osmer09}, but these SE mass scaling
relationships are calibrated to the RM-based \Hbeta\ mass scale rather
than by direct measurements of the time delays for these lines.  In this
way, SE masses can be estimated for large samples of AGNs observed in
spectroscopic surveys and across a large range of redshifts \citep[see,
e.g.,][]{Kollmeier06, YShen08, Shen11}.

This suite of intercalibrated AGN BH mass scaling relations initially
seems the ideal method for using the myriad of optical survey spectra of
AGNs to study BH-galaxy coevolution at all redshifts.  However, there
are significant concerns as to the reliability of masses based on lines
other than \Hbeta\ due to observed systematics in these masses compared
to \Hbeta\ masses.  For example, only tentative reverberation results
exist for \mgii\ \citep{Metzroth06}, so the use of \mgii\ as a virial
mass indicator is based only on its similar ionization potential and
line width compared to \Hbeta\ \citep{McLure04, Vestergaard&Osmer09}.
Studies have shown, however, that \mgii-based masses suffer from clear
systematics, some of which may be related to the Eddington ratio
\citep{Onken08} or systematic line width differences between \mgii\ and
\Hbeta\ \citep[see][]{Croom11}.  \mgii-based masses are also susceptible
to biases related to the prescription used for measuring the line width
\citep{Rafiee&Hall11}.

The same is true of SE masses based on \civ.  The current \civ\ mass
scaling relationship \citep[][hereafter VP06]{Vestergaard06} is based on
a direct calibration to \Hbeta\ reverberation-mapped AGNs.  For this
sample, VP06 found a relatively small scatter in the SE masses compared
with the RM masses, $\sim 0.3$ dex, which has since been cited as the
typical assumption of the accuracy of SE masses.  Additional studies
have also found a general consistency between \civ-based and
Balmer-line-based masses \citep[e.g.,][]{Greene10, Assef11}.  However,
other authors have questioned the reliability of \civ\ as a result of
finding little consistency or correlation and a large scatter between
\civ- and \Hbeta-based SE masses.  These studies contend that \civ-based
masses have too much scatter compared with \Hbeta\ masses to be a
reliable virial mass indicator \citep[e.g.,][]{Baskin&Laor05, Netzer07,
  Sulentic07, Shen12}.

One common postulate for this seeming unreliability is that the \civ\
emission region is susceptible to outflows and winds \citep[see][and
references therein]{Richards11}.  Such non-virial gas velocities are
then believed to bias the resulting SE mass measurements, rendering them
unreliable.  Interestingly, however, such dynamics do not seem to affect
the \civ\ RM-based masses.  In the few objects where RM measurements
have been made for multiple emission lines, \civ\ results follow the
expected virial relation with other low-ionization lines
\citep{Peterson00a} and the BH masses derived from the individual lines,
including \civ, are mutually consistent \citep{Peterson04}.  Here, we
explore this apparent contradiction --- why would RM-based \civ\ masses
behave as expected but SE masses do not?  In \S
\ref{S_SEmassAssumptions} we first review the fundamental assumptions
that make estimating a SE mass possible.  In \S \ref{S_RMmnrms} we
investigate properties of the \civ\ profile in the RM sample,
identifying a source of bias related to the \civ\ line profile.  In \S
\ref{S:shapebias} we quantify this bias and empirically fit a correction
to \civ\ masses that significantly reduces the scatter between \civ\ and
\Hbeta\ SE masses.  Finally, in \S\S \ref{S_discussion} and
\ref{S_conclusion} we put our results in context to other studies,
suggest a physical interpretation to explain our observations in terms
of an orientation-dependent outflow, and summarize our findings.

\section{Single-epoch Masses:  The Fundamental Assumptions}
\label{S_SEmassAssumptions}

We start by reviewing the assumptions made to estimate a SE BH mass.
These assumptions are based on the different means with which the
physical parameters required to measure a black hole mass (BLR radius
and velocity) are inferred in indirect (SE) and direct (RM) methods.

SE masses rely on the continuum luminosity as a proxy for the BLR
radius, and this is susceptible to two sources of uncertainty that do
not affect direct RM measurements of the BLR radius: variability and
host galaxy contamination.  First, since RM measures the BLR radius
directly from the delay between the continuum emission and the
reprocessed line emission, an RM mass is not directly dependent on the
luminosity.  Instead, the measured BLR radius and velocity (and
associated uncertainties) are physically connected to the intrinsic AGN
luminosity through photoionization and its effects on BLR emission
properties \citep[e.g., the locally optimally-emitting cloud
model;][]{Baldwin95}.  As a result, RM will always give the same mass
independent of the intrinsic variability \citep[cf.][]{Bentz07,
  Grier12b, Grier12a}.  However, the delay that makes RM experiments
possible means that the photons responsible for the luminosity measured
in a SE spectrum are not the same as those responsible for producing the
line emission whose width is measured in the same spectrum.  Therefore,
{\it the reliability of SE masses depends on the assumption that the
  difference between the observed SE luminosity and that responsible for
  triggering the observed line emission is negligible, or at least
  small.}  Fortunately, the variability amplitude of AGNs is relatively
small, $\sim 10-30$\%, on reverberation time scales --- the
light-crossing time of the BLR \citep[see, e.g.,][]{MacLeod10}.  Since
$M_{\rm SE} \propto L^{\sim 0.5}$, the uncertainty in the luminosity due
to variability is not a dominant source of uncertainty in SE masses and
is largely accounted for in the error budget through the uncertainties
in the calibration of the $R-L$ relationship itself
\citep[see][]{Bentz09rl}.  Second, only the AGN luminosity is correlated
with the radius of the BLR, since photoionization of the BLR gas from
the AGN continuum photons is what drives the $R-L$ relationship.
Therefore, SE masses will be overestimated if the observed SE luminosity
is not corrected for host galaxy starlight contamination.  This
contamination has been removed from RM data sets used to calibrate the
$R-L$ relationship \citep{Bentz06a, Bentz09rl}, but the RM masses
themselves are independent of this contamination.

Next, because velocity is as important a parameter as the luminosity in
determining SE BH masses \citep{Assef12} and $M \propto V^2$, we must
consider differences in the BLR velocity field characterizations of the
SE and RM methods.  RM masses are largely based on line widths measured
from the rms spectrum (i.e., the spectrum formed from the rms flux
deviations about the mean spectrum, based on all spectra obtained during
a single reverberation mapping campaign), so that only the kinematics of
the gas reverberating in response to the continuum photons contribute to
the mass determination.  The SE line profile is not necessarily
equivalent to the rms profile because there can be non-variable emission
components present.  So, {\it the reliability of SE masses also depends
  on the fundamental assumption that the SE line-width is an accurate
  proxy for the rms BLR velocity field}.  Comparisons of mean and rms
\Hbeta\ emission-line profiles show that even after removing the
contamination from narrow line region (NLR) emission, the SE profile is
not identical to the rms profile --- not all the \Hbeta\ broad
line-emitting gas is reverberating, particularly in the high-velocity
wings.  This results, on average, in narrower rms spectrum profiles than
mean spectrum profiles.  For \Hbeta, these differences are not
significant enough to strongly bias the SE mass estimates once the
differences are taken into account statistically by the calibration of
the scale factor, $\langle f \rangle$, for SE masses \citep[see,
e.g.,][]{Collin06, Park11}.

\section{The SE versus rms C\,{\scriptsize IV}  Profile}
\label{S_RMmnrms}

The many studies comparing SE \Hbeta\ masses to RM masses reinforces the
reliability of these masses, but no similar investigation exists for the
other emission lines used for SE masses.  The implication of (1) the
large scatter between \civ\ and \Hbeta\ SE masses and (2) the
consistency of \civ\ and \Hbeta\ RM masses strongly suggests that the
problem lies in the assumption that the \civ\ mean and rms line widths
are similar.

To examine this, we collected mean and rms line width characterizations
from RM experiments of both \Hbeta\ and \civ.  We take the \Hbeta\ FWHM
and line dispersion, $\sigma_l$ (the second moment of the line profile),
measurements from Table 1 of \citet{Collin06}, which have all been
measured homogeneously using the methods described by
\citet{Peterson04}.  Such a resource is not available for the \civ\ RM
data sets, so we acquired the original mean and rms spectra from all
available \civ\ RM experiments \citep{Clavel91, Reichert94, Korista95,
  Rodriguezpascual97, Wanders97, OBrien98, Peterson05, Metzroth06,
  Kaspi07}.  From these mean and rms spectra, we fit and subtracted a
linearly interpolated continuum between restframe continuum regions near
1450\AA\ (except for the NGC\, 7469 rms spectrum, which was fit near
1500\AA, where the continuum appeared lower) and 1700\AA.  We then
measured the \civ\ line widths directly (i.e., no functional forms were
fit), except for the case of the rms spectrum of NGC\, 4395, which was
too noisy to measure directly.  In this case, a sixth order
Gauss-Hermite polynomial was fit to the data, and the width was measured
from the fit.  We employ the same line width measurement techniques as
those utilized for \Hbeta\ in order to have a homogeneous comparison.
These new \civ\ measurements are given in Table 1.

\begin{deluxetable*}{llccccc}
\tablecolumns{7}
\tablewidth{400pt}
\tablecaption{\civ\ Emission Line Widths of Reverberation Sample}
\tablehead{
\colhead{}&\colhead{}&\multicolumn{2}{c}{mean spectrum}&\multicolumn{2}{c}{rms spectrum}&\colhead{}\\
\hline
\colhead{Object}&\colhead{Data Set\tablenotemark{1}}&\colhead{FWHM}&\colhead{$\sigma_l$}&\colhead{FWHM}
&\colhead{$\sigma_l$}&\colhead{Ref.\tablenotemark{2}}}

\startdata
3c390.3    & \nodata  &  6000$\pm$150 & 5360$\pm$40 & 11340$\pm$700 &   5760$\pm$ 50 & 1 \\
Fairall\,9 & \nodata  &  3040$\pm$ 50 & 4100$\pm$40 &  3290$\pm$900 &   2670$\pm$130 & 2 \\
NGC\,3783  & \nodata  &  3120$\pm$ 60 & 2910$\pm$20 &  4060$\pm$200 &   2790$\pm$ 70 & 3 \\
NGC\,4151  & 1988     &  3180$\pm$130 & 5190$\pm$30 & 13820$\pm$530 &   5740$\pm$ 70 & 4 \\
NGC\,4151  & 1991     &  6610$\pm$260 & 4800$\pm$30 & 10450$\pm$280 &   5160$\pm$ 90 & 4 \\
NGC\,4395  & Visit 2  &  1070$\pm$ 10 & 2130$\pm$20 &  5840$\pm$350 &   2510$\pm$ 50 & 5 \\
NGC\,4395  & Visit 3  &  1270$\pm$ 10 & 1950$\pm$20 &  1900$\pm$170 &    990$\pm$150 & 5 \\
NGC\,5548  & {\it HST}&  4080$\pm$ 70 & 3940$\pm$10 &  6840$\pm$150 &   3640$\pm$ 30 & 6 \\
NGC\,5548  & {\it IUE}&  4860$\pm$250 & 4100$\pm$60 &  5590$\pm$200 &   4110$\pm$ 60 & 7 \\
NGC\,7469  & \nodata  &  3290$\pm$ 90 & 3570$\pm$30 &  4380$\pm$190 &   2670$\pm$ 30 & 8 \\
S50836+071\tablenotemark{3} & \nodata  &  8920$\pm$190 & 4600$\pm$30 &    \nodata    &     \nodata    & 9
\enddata

\tablenotetext{1}{For objects with multiple measurements, we include a
  note to distinguish between them; see References for details.}
\tablenotetext{2}{Original Data References: (1) \citet{OBrien98}; (2)
  \citet{Rodriguezpascual97}; (3) \citet{Reichert94}; (4)
  \citet{Metzroth06}; (5) \citet{Peterson05}; (6) \citet{Korista95}; (7)
  \citet{Clavel91}; (8) \citet{Wanders97}; (9) \citet{Kaspi07}.}
\tablenotetext{3}{Unfortunately, we had to exclude this object from our
  analysis because the rms spectrum was very noisy, and the \civ\ variable
  emission was blended with variable \heii\,$\lambda$\,1640, so a
  reliable \civ\ line width could not be measured from the rms spectrum.}

\label{Tab_Widths}
\end{deluxetable*}

\begin{figure}
\figurenum{1}
\epsscale{1.0}
\plotone{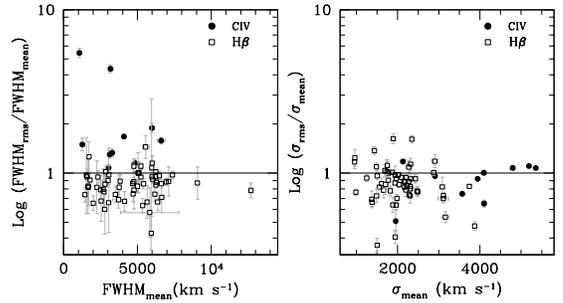}

\caption{\civ\ and \Hbeta\ line width ratios measured from the mean and
  rms spectra from reverberation mapping experiments.  The left and
  right panels show ratios of the widths measured using the FWHM and
  $\sigma_l$, respectively.  Solid (open) points represent measurements
  of \civ\ (\Hbeta) emission-line widths.  The solid horizontal lines
  corresponds to a ratio of unity.}

\label{Fig:widthmnvsrms}
\end{figure}

Figure \ref{Fig:widthmnvsrms} shows the ratios of the mean and rms line
widths as a function of the line width for \Hbeta\ (open symbols) and
\civ\ (closed symbols).  We consider first the results for the FWHM.
The known trend that \Hbeta\ shows systematically narrower rms than mean
line widths \citep{Collin06} is clearly apparent, as the \Hbeta\ ratios
lie mostly below unity.  On the other hand, \civ\ shows a systematically
different trend, with all objects having {\it broader} rms than mean
spectrum line widths, and in some cases the difference is quite large.
The line dispersion ratios for \Hbeta\ show a similar offset from unity,
albeit with greater scatter.  On the other hand, the \civ\ line
dispersion ratios are now uniformly scattered around unity. Figure
\ref{Fig:widthmnvsrms} shows that there are significant differences
between the time variable and instantaneous or mean characterization of
the \Hbeta\ and \civ\ broad emission-line widths.  These differences are
larger for the FWHM than for the line dispersion, and they are the
largest and least uniform for the FWHM characterization of the \civ\
line.

\begin{figure}
\figurenum{2}
\epsscale{1.0}
\plotone{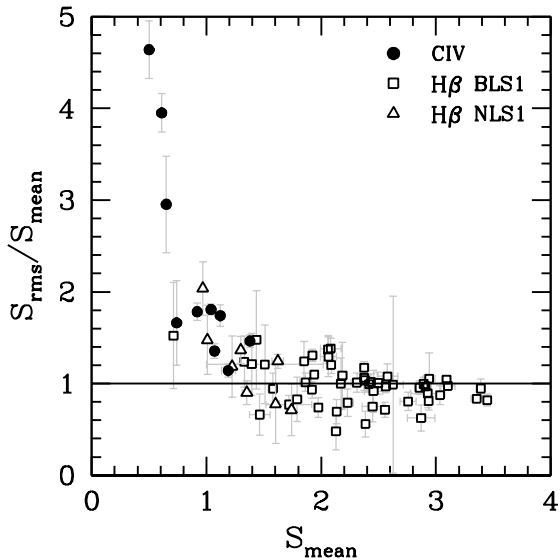}

\caption{\civ\ and \Hbeta\ line shape ratios measured from the mean and
  rms spectra from reverberation mapping experiments. Solid points
  represent measurements of \civ\ emission-line widths, open points
  refer to \Hbeta\ widths.  The \Hbeta\ measurements have been divided
  by symbol type into those from normal broad-line AGNs (open squares)
  and NLS1s (open triangles). The solid horizontal line represents the
  1:1 relation of equal mean and rms shapes.  (There are no \civ\ points
  with $S_{\rm mean} > 1.4$).}

\label{Fig:shaperesidvsshape}
\end{figure}

We investigate this incompatibility further by parameterizing the shape
of the line profile in terms of the ratio $S = {\rm FWHM}/\sigma_l$.
Profiles with lower values of $S$ have `peaky' profiles with larger
kurtosis, while those with higher values of $S$ have `boxy' profiles
with smaller kurtosis.  A Gaussian profile has $S=2.35$ and a kurtosis
of three.  Figure \ref{Fig:shaperesidvsshape} shows the ratio between
the line shapes determined from the mean and rms spectra, $S_{\rm
  rms}/S_{\rm mean}$, for both \Hbeta\ and \civ.  The shapes of the mean
and rms \Hbeta\ line profiles are essentially indistinguishable.  There
may be a weak effect associated with narrow line Seyfert 1's (open
triangles), where it is difficult to accurately remove the contribution
from the narrow-line component.  For \civ, however, the shapes of the
mean and rms line profiles tend to be very different.  The consistency
in the mean and rms \Hbeta\ line shape allows the simple offset in
\Hbeta\ widths seen in Figure \ref{Fig:widthmnvsrms} to be easily
compensated for in the standard recalibration of the RM BH masses for SE
data.  No such calibration would work for \civ\ without introducing
extra scatter in the SE mass estimates compared to RM mass estimates.

The differences between the SE and rms \civ\ profile can also be seen
through visual inspection of the mean and rms spectra observed in \civ\
reverberation mapping campaigns.  Figure \ref{Fig:civmnrms} shows the
mean and rms profiles for the RM sample, excluding the $z \sim 2$ quasar
monitored by \citet{Kaspi07}.  The black curves show the mean spectrum
\civ\ profile, normalized to the peak flux after subtracting a local,
linearly interpolated continuum, and the gray, dashed-dotted curves show
the corresponding continuum-subtracted normalized \civ\ line profile
from the rms spectrum.  Here the continuum was fit to both the mean and
rms spectra using regions near restframe 1450\AA\ and 1700\AA.
Normalizing in this way makes it easy to see that the FWHM in the rms
spectrum is broader than that in the mean spectrum for all objects, as
was also seen in Figure \ref{Fig:widthmnvsrms}.  More interestingly,
however, are the visibly different shapes of the \civ\ lines in the mean
and rms spectra.  This was also already demonstrated in Figure
\ref{Fig:shaperesidvsshape}, but we see in Figure \ref{Fig:civmnrms}
that the mean spectrum has an additional, low-velocity component absent
from the rms spectrum for all objects but Fairall 9.  In the case of
Fairall 9, the rms signal is weak and the strength of the narrow core
component in the rms spectrum is within the level of the flux
uncertainties of the IUE campaign data, which makes it difficult to
interpret the shape differences \citep[see][]{Rodriguezpascual97}.  In
general, it appears that the SE or mean \civ\ profile is made up of two
components: a non-variable, largely core, component plus a variable
component, the velocity profile of which can be isolated in the rms
spectrum.  We can emphasize which portion of the mean spectrum is
variable with the solid gray curves in Figure \ref{Fig:civmnrms} that
show the normalized rms spectra scaled in flux by an arbitrary fraction
to match (by eye) the mean and rms spectrum flux level in the red wing
of \civ.

\section{Shape Bias in  C\,{\scriptsize IV}  BH Masses}
\label{S:shapebias}

The existence of two components in the SE \civ\ profiles, only one of
which is variable, has significant implications for the determinations
of SE \civ\ BH masses, both for individual estimates but also for
calibrating the \civ\ mass scale itself.  The contribution of the
non-variable component creates a bias to the relevant (i.e.,
reverberating) BLR velocity field.  This is particularly true when using
the FWHM to characterize the line width.  Furthermore, the strength and
profile of the non-variable component is not uniform object-to-object,
so a simple change in the scale factor, $\langle f \rangle$, or
subtraction of a constant, narrow velocity component will not work.
Instead, it seems likely that any correction must be based on the shape
of the individual \civ\ line profile.  

To explore this, we compiled several samples from the literature that
(1) have both \civ\ and \Hbeta\ mass estimates, and (2) for which both
FWHM and $\sigma_l$ characterizations of the line width are published or
for which we can measure them from the original \civ\ spectra so that we
can calculate the shape parameter, $S={\rm FWHM}/\sigma_l$.  These samples
include those of VP06, \citet[][hereafter A11]{Assef11}, and
\citet[][hereafter N07]{Netzer07}.  A11 provide all the needed
quantities in their Tables 3 and 4.  We use their ``Prescription A''
because it is most similar to the method used by VP06.  We also consider
both the group I and group II objects from A11, excluding only the
broad absorption line quasar, H1413+117, leaving a sample of 11 objects.

\begin{figure}
\figurenum{3}
\epsscale{1.0}
\plotone{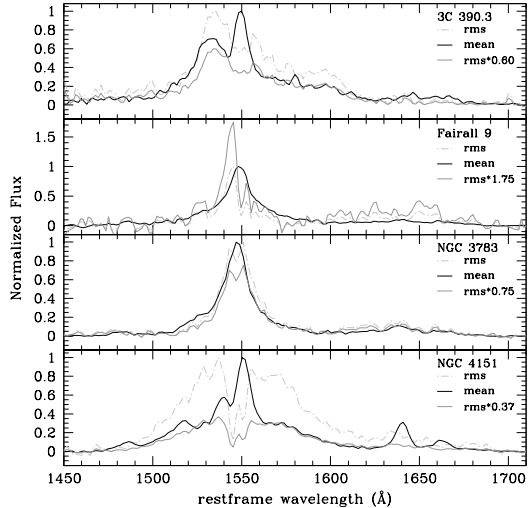}

\caption{Mean (black) and rms (gray dot-dashed) spectra of the \civ\
  reverberation mapping sample. Spectra have been continuum subtracted
  and normalized in flux based on the \civ\ emission-line peak. Solid
  gray curves show the rms spectra scaled by an arbitrary factor to
  approximately match the red wing flux between each rms and mean
  spectrum (see Section \ref{S_RMmnrms} for details).  The red curves in
  the panel for NGC\, 4395 show the normalized (red dot-dashed) and
  scaled (red solid) Gauss-Hermite polynomial fit to the rms spectrum
  used for line width measurements.  We show only one set of spectra for
  each object, even though some objects were monitored more than once;
  results not shown are qualitatively similar to those here.}

\label{Fig:civmnrms}
\end{figure}

\begin{figure}
\figurenum{3}
\epsscale{1.0}
\plotone{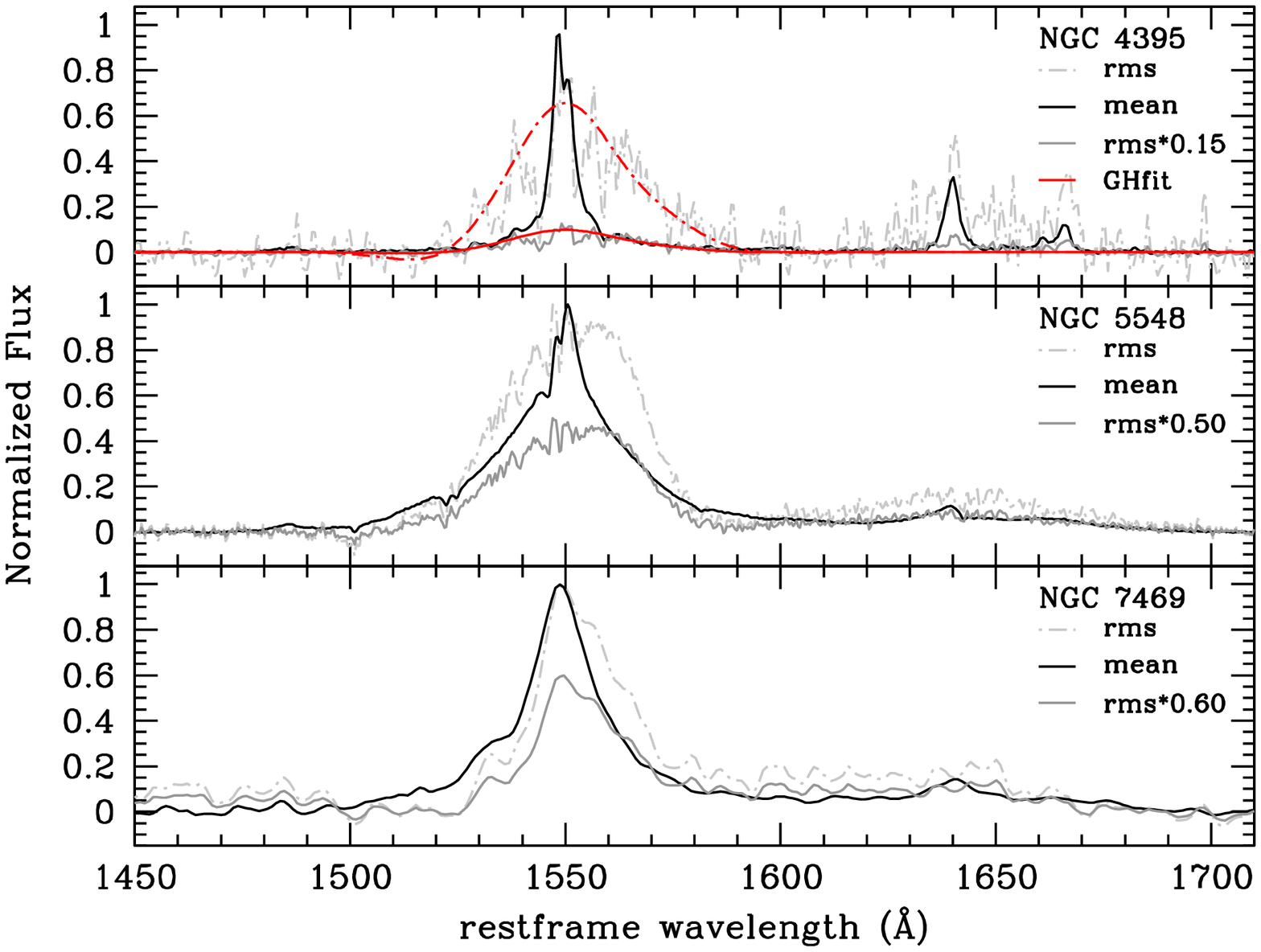}

\caption{{\it Continued.}}

\end{figure}

The VP06 sample is of particular interest for investigating biases,
since it calibrates the \civ\ SE BH mass scale.  VP06 often used
multiple SE spectra of a single object and calculated masses from each
epoch.  Furthermore, the epochs and number of spectra they used varied
from object to object and between \civ\ and \Hbeta.  Here we are more
interested in the SE \civ\ and \Hbeta\ masses and $S$ for different
objects rather than different epochs, so we derived a geometric average
of each of these quantities for the VP06 sample.  For objects with more
than one epoch of data, we use the \civ\ FWHM and $\sigma_l$ to
calculate $\log(S)$ for each epoch, and then calculate a simple average
of the individual values of $\log(S)$, adding uncertainties from each
epoch in quadrature\footnote{Intrinsic variations in $S$ could occur
  from epoch to epoch, so averaging over epochs may be removing
  information about long timescale changes in emission-line profiles
  known to exist.  However, the standard deviation about the mean $S$ in
  all objects for which we calculate an average is much smaller than the
  formal uncertainties in $S$ propagated from the uncertainties in the
  measurements of the SE FWHM and $\sigma_l$.}.  We also average the SE
\civ\ masses and uncertainties given in Table 2 of VP06 for each object
in the same way.  Finally, we recalculate the SE \Hbeta\ mass using the
updated formula provided by A11 (their Equation 4), which is based on
the most recent calibration of the \Hbeta\ $R-L$ relation of
\citet{Bentz09rl}.  We then similarly average these SE masses to end up
with one \Hbeta\ mass for each object.  Objects that had SE \Hbeta\
masses but not SE \civ\ masses were omitted from the sample.  For the
six objects lacking an SE \Hbeta\ mass, we used the RM mass, which is
generally equivalent\footnote{This is not strictly correct because host
  galaxy starlight has not been removed from the SE luminosities.  The
  resulting \Hbeta\ masses for these six objects will be somewhat
  underestimated compared to the expected SE \Hbeta\ mass (this
  underestimation is likely $\lesssim 0.2$ dex, based on the average
  difference calculated from the rest of the sample, for which both SE
  and RM masses are available).}.  This results in a sample of 27
objects taken from VP06.

\begin{figure}
\figurenum{4}
\epsscale{1.0}
\plotone{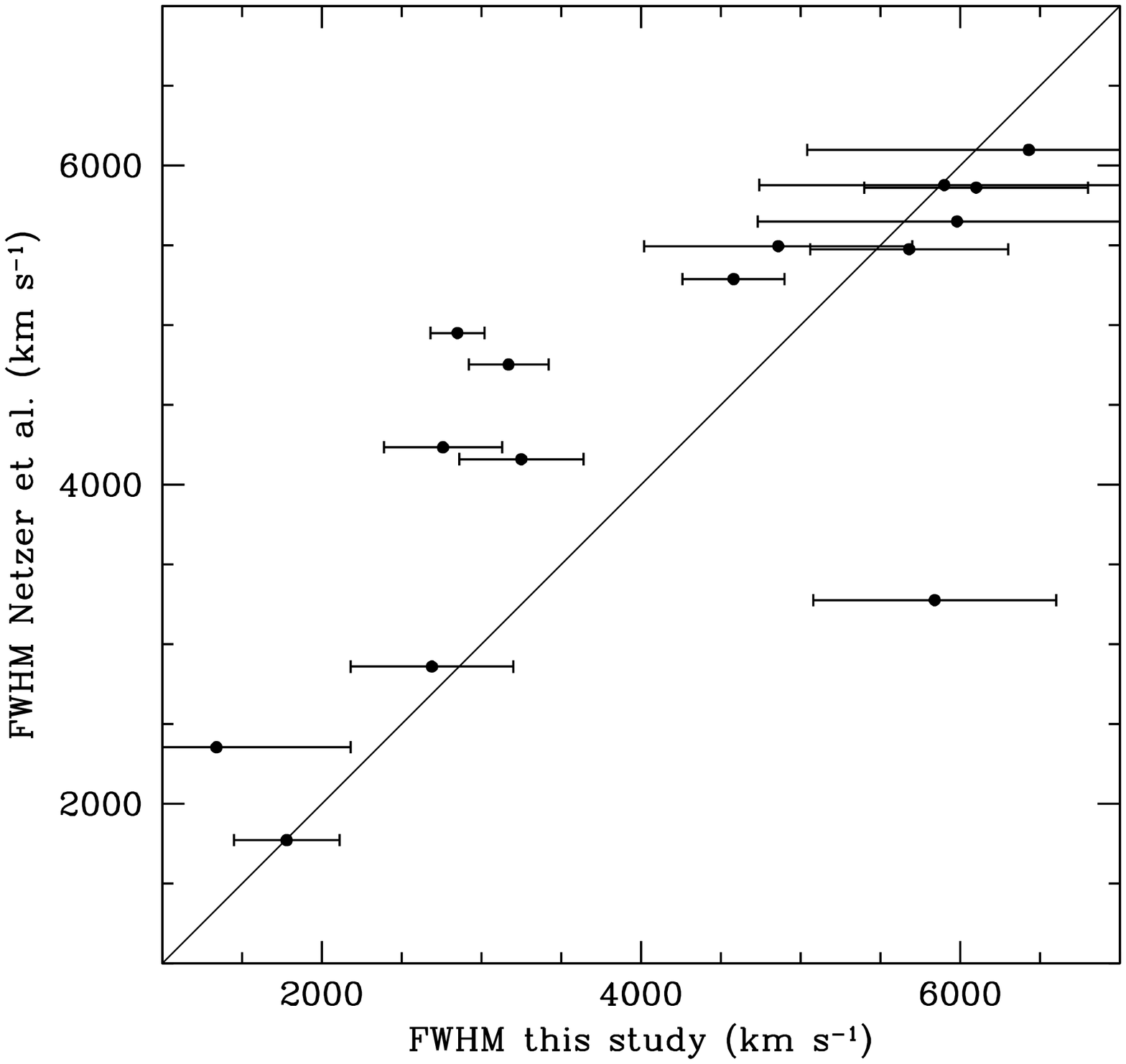}

\caption{Comparison of \civ\ FWHM measurements taken from N07 for their
  sample of 15 SDSS AGNs and those remeasured in this study.  Error bars
  are only shown for our new measurements because N07 do not provide
  uncertainties on their width measurements.  The solid line shows the
  expected 1:1 relation in width.}

\label{Fig:usVnetzerFWHM}
\end{figure}

N07 did not publish $\sigma_l$ values for their sample, but their study
was based on SDSS spectra. Thus, we obtained the SDSS spectra and
remeasured both the FWHM, $\sigma_l$, and their uncertainties following
the methods for ``Prescription A'' described by A11.  This includes
fitting sixth order Gauss-Hermite polynomials to the \civ\ profiles
because of the typically poor data quality of this sample (average S/N
per pixel in the continuum near restframe 1700\AA\ is only 5.4).
Interestingly, in several cases, our measurement for the FWHM of the
\civ\ line in the N07 sample differs significantly from the values
presented by N07, as shown in Figure \ref{Fig:usVnetzerFWHM}.
Unfortunately, N07 neither describe how they determined the widths of
the lines given in their Table 2 nor provide any uncertainties on these
measured values, so reconciling these differences is not possible here.
Nonetheless, this highlights the difficulties in these types of studies
as different measurement techniques applied to the same data set can
result in statistically different measurements of the same quantities
\citep[see also][]{Assef11, Vestergaard11, Park11}.  We will continue to
use our own FWHM measurements of the N07 sample because they have been
determined with methods consistent with those applied to the other
samples used here.  We flag seven of these objects not necessarily
because they are all outliers in any part of our analysis, but because
we have evaluated their line widths to be potentially unreliable due to
evidence for absorption in the \civ\ line profiles of the SDSS
spectra\footnote{The classification of absorption {\it here} is based
  solely on the observed SDSS spectra.  However, the existence of some
  such absorption has been verified in new, higher $S/N$ data obtained
  for seven of the N07 sources \citep[][Denney et al.\ in
  prep]{Assef11}.}.

Using our \civ\ line widths and the continuum luminosities given by N07,
we calculate new SE \civ\ masses for each of the 15 objects in this
sample based on the FWHM and $\sigma_l$ following Equations 7 and 8 of
VP06 (also Equation 6 of A11).  We also recalculate the SE \Hbeta\ mass
using the FWHM and luminosity provided by N07 but using the formula
given by A11.  Unfortunately, we could not calculate uncertainties in
the mass estimates because N07 do not provide uncertainties on the
\Hbeta\ width or optical/UV luminosity measurements.  Nonetheless, this
partial reanalysis places all masses on the same mass scale, with
homogeneously measured \civ\ line widths.

\begin{figure}
\figurenum{5}
\epsscale{1.0}
\plotone{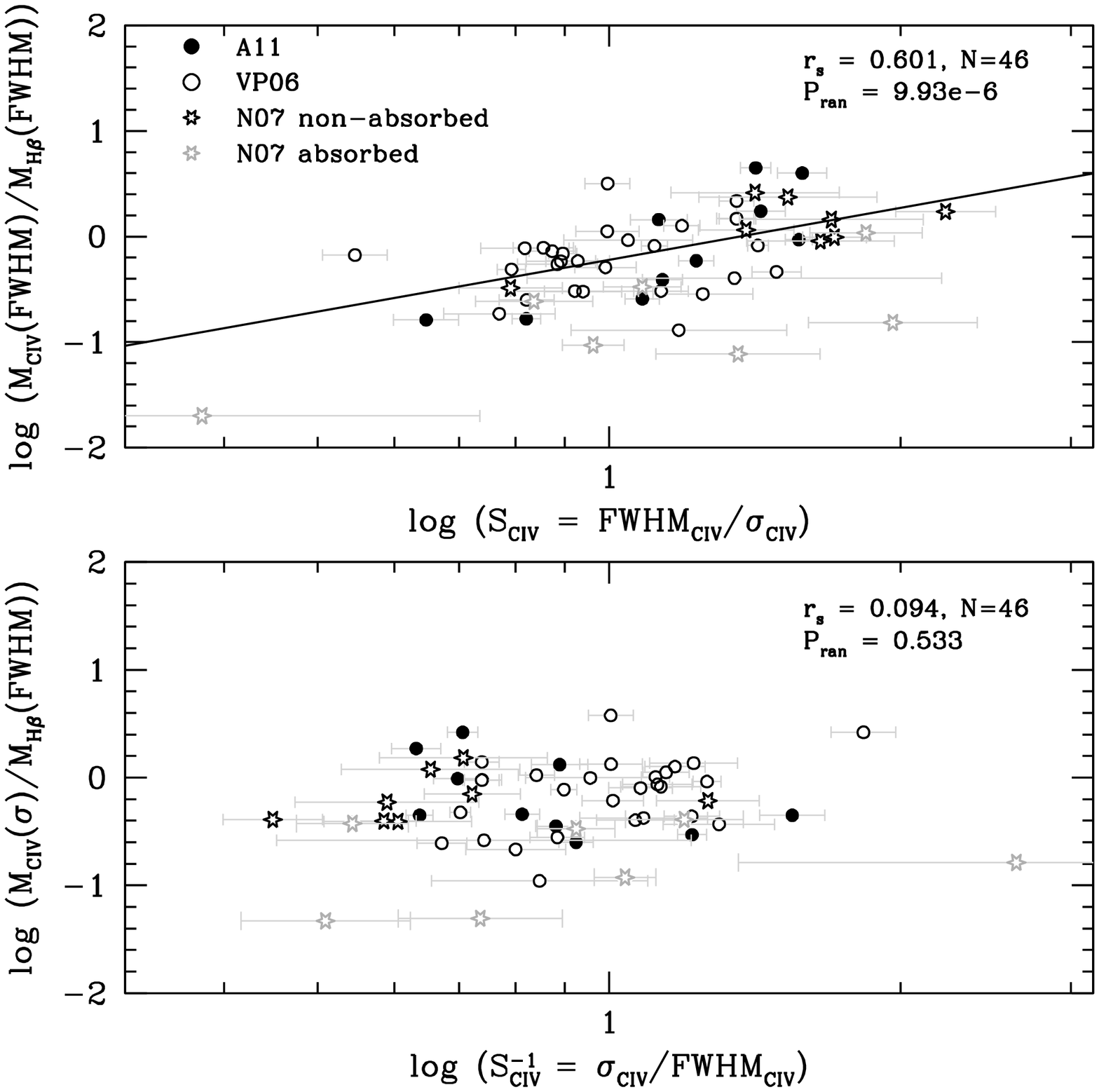}

\caption{{\it Top}: Comparison between \civ\ and \Hbeta\ mass residuals,
  calculated using FWHM, and the shape of the \civ\ line, $S={\rm
    FWHM}/\sigma_l$ for the A11 (solid circles), VP06 (open circles),
  and N07 (stars) samples.  The solid line shows the best fit
  correlation to the combined sample, excluding absorbed N07 objects
  (gray stars). The Spearman rank order coefficient, $r_{\rm s}$, the
  number of objects included in the fit, $N$, and the probability that
  the correlation is found by chance, $P_{\rm ran}$, are given in the
  top right corner.  Error bars are not included on the mass residuals
  because uncertainties could not be calculated similarly for all
  samples (see Section \ref{S:shapebias}).  {\it Bottom}: Same as top
  panel except the \civ\ masses were calculated using $\sigma_l$ and the
  resulting residuals with the \Hbeta\ masses are compared to $S^{-1}$
  so that both plots have the same dependence on the \civ\ width
  characterization used in the mass estimates. Note: $r_{\rm s}$ is
  insensitive to the choice to use $S$ or $S^{-1}$ as the independent
  variable.}

\label{Fig:massresidVshape}
\end{figure}

In Figure \ref{Fig:massresidVshape}, we show the ratio of \civ\ to
\Hbeta\ SE masses as a function of the line shape, $S$.  The top panel
shows the BH masses calculated using the FWHM of the line.  The
different symbol types represent the different samples.  There is a
statistically significant correlation of the mass residuals with \civ\
line shape.  The Spearman rank order coefficient, given in the top right
corner of the plot, is $r_{\rm s}=0.601$, and the probability that such
a correlation would arise by chance if the quantities were intrinsically
uncorrelated is $P_{\rm ran}= 9.93 \times 10^{-6}$. These statistics
were determined with only 46 of the 53 original objects because we
omitted the seven absorbed N07 targets (gray stars) from the analysis.

An initial concern when correlating the line shape, which is determined
from the \civ\ FWHM, with the BH mass residuals, which are also a
function of the \civ\ FWHM, is that the correlation is a consequence of
this redundant dependence, effectively plotting FWHM vs.\ FWHM.  To test
this possibility, we also calculated the \civ\ BH mass using $\sigma_l$
and compare the resulting mass residuals with \Hbeta\ to the inverted
the line shape parameter, $S^{-1}$, in the bottom panel of Figure
\ref{Fig:massresidVshape}.  The \Hbeta\ SE BH mass is still determined
using the FWHM.  Now both axes depend on $\sigma_{\civm}$ in the same
way they depend on FWHM$_{\civm}$ in the top panel.  If the correlation
in the top panel was due only to a redundant dependence on line width in
both coordinate axes, we would expect to see a correlation of similar
strength in the bottom panel, but we find no correlation, with $r_{r\rm
  s}=0.094$ and $P_{\rm ran}= 0.533$.  Furthermore, a partial
correlation test between the quantities correlated in the top panel,
with FWHM$_{\civm}$ as a nuisance variable, results in a partial
correlation coefficient, $r=0.389$, which is nearly equal to that from
the equivalent test for the bottom panel, with $\sigma_{\civm}$ as the
nuisance parameter, i.e., $r=0.381$, yet again, we don't see a
correlation of the $\sigma_{\civm}$-based masses with $S^{-1}$.  This
not only confirms that the strong correlation of FWHM-based masses with
shape is real (i.e., not due to a redundant dependence of BH mass and
shape on line width), but also that \civ\ masses based on $\sigma_l$
more reliably reproduce, in general, the \Hbeta\ mass estimates.  This
latter point is not altogether surprising, since $\sigma_l$ depends less
on the structure of the core of the line profile, where the non-variable
\civ\ emission component most strongly affects the shape.
 
\begin{figure}
\figurenum{6}
\epsscale{1.0}
\plotone{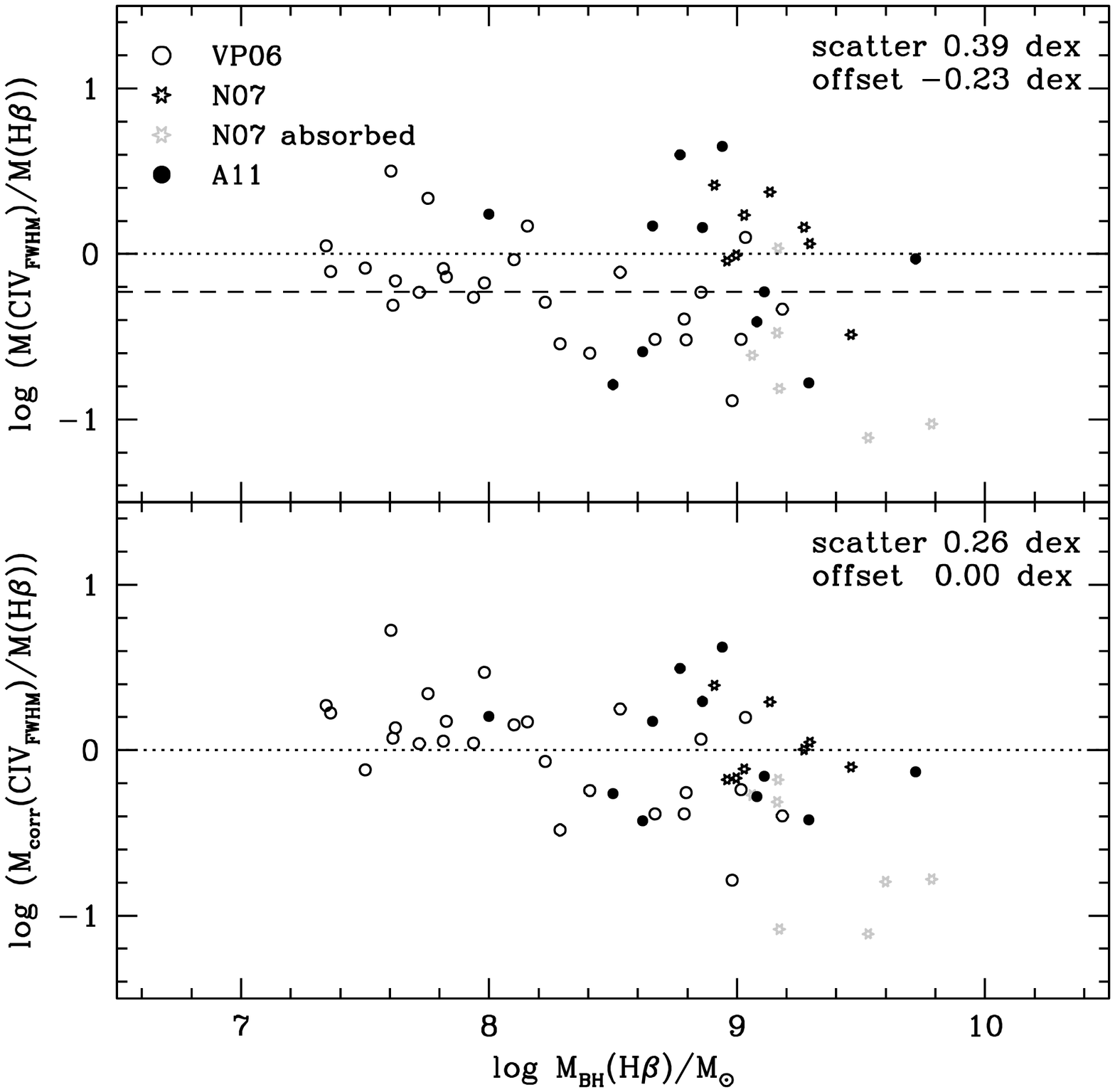}

\caption{\civ-to-\Hbeta\ SE BH mass residuals from the VP06, A11, and N07
  samples as a function of \Hbeta\ mass.  Symbol types are the same as
  Figure \ref{Fig:massresidVshape}. The top panel shows residuals
  calculated with the original \civ\ masses.  The bottom panel shows the
  residuals after correcting the \civ\ masses using Equation (1).  The
  mean scatter about and offset from (dashed line) the 1:1 relation
  (dotted line) for the combined sample is given in the top corner of
  each panel (The corrected scatter and offset exclude the N07 absorbed
  objects.).}

\label{Fig:origAcorrmassresid}
\end{figure}

The strength of the correlation between the FWHM-based mass residuals
and \civ\ line shape suggests that this shape trend could explain a
significant fraction of the scatter between SE \civ\ and \Hbeta\ masses
observed in the literature, as the studies discussed in the Section
\ref{S_Intro} generally used FWHM-based \civ\ masses.  We can fit the
correlation of the mass differences with shape to derive a corrected
FWHM-based \civ\ SE mass equation of the form

\begin{align}
  \log M_{\civm}^{\rm corr}({\rm FWHM}) = \log M_{\civm}^{\rm orig}({\rm FWHM}) + 0.219 \nonumber \\
 - 1.63 \log \left(\frac{{\rm FWHM}_{\civm}}{\sigma_{\civm}}\right) ,
\end{align}

\noindent where the uncertainties in the slope and the intercept are
given in Table \ref{Tab_Fits}.  The top panel of Figure
\ref{Fig:origAcorrmassresid} shows the original FWHM-based
\civ-to-\Hbeta\ mass residuals as a function of \Hbeta\ SE BH mass.  The
scatter in the combined sample about the 1:1 relation of equal
\civ-to-\Hbeta\ mass is 0.39 dex, and there is a mean offset of $-0.23$
dex.  The bottom panel shows the same residuals after correcting for the
\civ\ line shape bias using Equation (1).  There is no longer a mean
offset and the scatter is now only 0.26 dex.  We have excluded the
absorbed N07 targets (gray stars) from the fit and latter scatter
determination because the \civ\ mass of these objects is biased by
factors other than the intrinsic \civ\ line shape (e.g., absorption; see
also Denney et al.\ in prep.).

\begin{deluxetable}{lccc}
\tablecolumns{4}
\tablewidth{0pt}
\tablecaption{Fit Parameters to Correlation between BH Mass Residuals and \civ\ Line Shape}
\tablehead{
\colhead{Data Set}&\colhead{Slope}&\colhead{Intercept}&\colhead{Scatter}}

\startdata
A11       &  3.65  $\pm$ 0.79  &  -0.336 $\pm$ 0.101 &  0.20 dex\\
N07       &  1.49  $\pm$ 0.67  &  -0.172 $\pm$ 0.143 &  0.16 dex\\
VP06      &  2.01  $\pm$ 0.64  &  -0.046 $\pm$ 0.062 &  0.26 dex\\
Combined  &  1.63  $\pm$ 0.38  &  -0.219 $\pm$ 0.051 &  0.26 dex
\enddata

\label{Tab_Fits}
\end{deluxetable}

Despite the decreased scatter, there are still outliers.  Many of these
are the previously discussed N07 objects.  The largest A11 sample
outlier is Q0142$-$100, but this object was classified as a group II
object by A11, due to an unreliable \Hbeta\ line width, so the problem
may have nothing to do with \civ.  The two largest VP06 outliers (one
above and one below the 1:1 relation) are the two radio loud AGNs in the
RM sample, 3C\,120 and 3C\,390.3.  This may simply be a coincidence or
may indicate that radio loud AGNs follow a different \civ\ shape
relation than radio quiet AGNs.  For example, \citet{Richards11} found
very distinct differences between the \civ\ properties of the radio-loud
and radio-quiet populations.

\section{Discussion}
\label{S_discussion}

\subsection{Other Line Width and Shape Characterizations}

The FWHM is the most widely-used line-width characterization in the
literature, and it has the advantages of being easy to measure and
relatively insensitive to blending in the line wings.  On the other
hand, measurements of $\sigma_l$ are robust for a wide range of line
profiles and have been found to be a less biased line characterization
in some cases \citep[see, e.g.,][]{Peterson04, Denney09a,
  Rafiee&Hall11}.  The problems with measuring $\sigma_l$ include
difficulties in accurately defining the line wings, particularly in low
$S/N$ spectra, and in the presence of other blended emission-line
features.  Despite these concerns, Figures \ref{Fig:widthmnvsrms} and
\ref{Fig:massresidVshape}, respectively, show that characterizing the
\civ\ width with $\sigma_l$ leads to (1) velocities that are more
consistent with the 1:1 relation between the mean and rms spectrum in
the reverberation-mapped sample, and (2) masses that show less bias with
the line shape, $S$. In fact, when we calculate the \civ\ mass using
$\sigma_l$ (i.e., Equation 8 of VP06), the scatter in the mass residuals
about the 1:1 relation of 0.28 dex is only slightly larger than the
scatter in the corrected FWHM-based masses (0.26 dex) shown in Figure
\ref{Fig:origAcorrmassresid}, albeit with an additional systematic
offset of $-0.17$ dex indicating a zero-point calibration difference
between the full sample and VP06.  This offset could be due to the \civ\
shape bias or other systematic uncertainties in the exact prescription
for calculating $\sigma_l$, such as blending with the red shelf
\citep[see, e.g.][]{Fine10, Assef11}.  In either case, if we adjust the
zero-point of the VP06 relation to eliminate the offset, the remaining
scatter about unity is only 0.26 dex, equivalent to the FWHM-corrected
\civ\ masses.

Because our shape correction to the \civ\ mass is a combination of only
the FWHM and $\sigma_l$, expanding Equation (1) shows that we have
effectively fit a virial relation of the form $M \propto {\rm FWHM}^x
\sigma_l^{y}$, where $y=2-x$ in general, and here, $y=1.63$ and
$x=0.37$.  The preferentially higher weight of the line dispersion in
this expansion explains the similarity of the scatter between the
shape-corrected masses and $\sigma_l$-based masses.  We also determined
the \civ\ shape correction strictly from the kurtosis of the \civ\
profile.  Results are not shown because they were only based on the N07
and A11 samples for which we had access to the data, but are consistent
with using $S = {\rm FWHM}/\sigma_l$.  This demonstrates that utilizing
either (1) a combination of line width plus shape, or (2) a more complex
(i.e., than FWHM) characterization of the \civ\ line profile is
necessary to properly calibrate \civ\ and \Hbeta\ masses due to the
contamination of the non-variable component of \civ.
\citet{Baskin&Laor05} also found that the scatter between \civ\ and
\Hbeta\ masses could be reduced by fitting a \civ\ mass correction based
on additional \civ\ profile information, albeit with more parameters, to
what we have done here.  Additionally, \citet{Wang09} and \citet{Park11}
have suggested relaxing a strict virial requirement ($M \propto V^2$)
altogether as a means to reduce scatter in the calibration of SE mass
scales due to systematic differences between various SE line profiles
and the rms \Hbeta\ profile to which these mass scales are ultimately
calibrated.  \citet{Rafiee&Hall11} found that this relaxation could also
explain the ``sub-Eddington boundary'' for FWHM-based BH masses, but
then again, so could using $\sigma_l$-based masses with the typical
virial requirement.

\subsection{Sources of Remaining Scatter}
\label{S:remainingscatter}

While the \civ\ shape correction defined by Equation (1) significantly
reduces (by 0.13 dex) the scatter in the \civ-to-\Hbeta\ mass residuals,
a scatter of $\sim$0.3 dex still remains.  We briefly investigated three
potential sources for this remaining scatter:
\begin{enumerate}

\item {\bf Data quality:} Several absorbed N07 objects (gray stars)
  became even larger outliers after the \civ\ shape correction.  This
  suggests that the {\it measured} line width in these objects is not
  the {\it intrinsic} line width, so the correction to these sources was
  meaningless.  This is a serious concern for SE masses measured from
  survey quality spectra, which are typically of relatively low $S/N$
  \citep[see][and Denney et al., in prep.\ for additional discussions
  and investigations of this concern]{Denney09a, Assef11}.

\item {\bf Data Inhomogeneity:} As there is no universal prescription
  for AGN spectral decomposition and line width measurements,
  inhomogeneity in the data analysis and computational methods used to
  derive SE masses leads to measurable systematics when trying to
  compare masses between samples and emission lines.  This is
  demonstrated by the comparison of our FWHM measurements of the N07
  sample in Figure \ref{Fig:usVnetzerFWHM} \citep[see also][for
  additional sample to sample comparisons]{Assef11, Vestergaard11}.
  There are also no universal mass scaling relationships for each
  emission line, and differences in the mass scale also lead to
  differences in masses derived by different studies \citep[see
  discussions by][]{McGill08, Shen12}.

\item {\bf H\boldmath$\beta$ Systematics:} Literature comparisons
  between SE \civ\ and \Hbeta\ masses typically blame \civ\ masses for
  any observed inconsistencies, since \Hbeta\ is arguably the most
  well-characterized and best-studied line in terms of direct RM-based
  masses for comparison.  However, \Hbeta\ it is not without systematics
  of its own, particularly contamination by NLR emission and host galaxy
  starlight, which are relatively unimportant for \civ.  The systematics
  introduced into the \Hbeta\ line width and/or optical luminosity can
  be significant \citep{Denney09a, Bentz09rl, Park11}.
  
\end{enumerate}

We tested if the scatter in the \civ-to-\Hbeta\ masses could be further
reduced by addressing these additional sources of scatter.  First, we
removed (1) single VP06 epochs with $S/N < 10$ pixel$^{-1}$ in the
continuum, (2) the seven objects from the N07 sample that show evidence
for absorption (we could not cull $S/N < 10$ pixel$^{-1}$ data from this
sample without eliminating it completely), and (3) the three A11 sources
defined as Group II objects due to unreliable \Hbeta\ widths and other
issues \citep[see][for details]{Assef11}.

Next, we removed host starlight from the VP06 sample by replacing the SE
\Hbeta\ masses with the RM \Hbeta\ masses.  Ideally, host starlight
should be removed from all samples, but the required host fluxes are not
available.  In any case, the VP06 sample is the most susceptible to this
bias because of its relatively lower mean AGN luminosity.  This choice
removes some of the consistency between the treatment of this sample
compared to the A11 and N07 samples, but these differences are smaller
than the accuracy gained by this correction.

\begin{figure}
\figurenum{7}
\epsscale{1.0}
\plotone{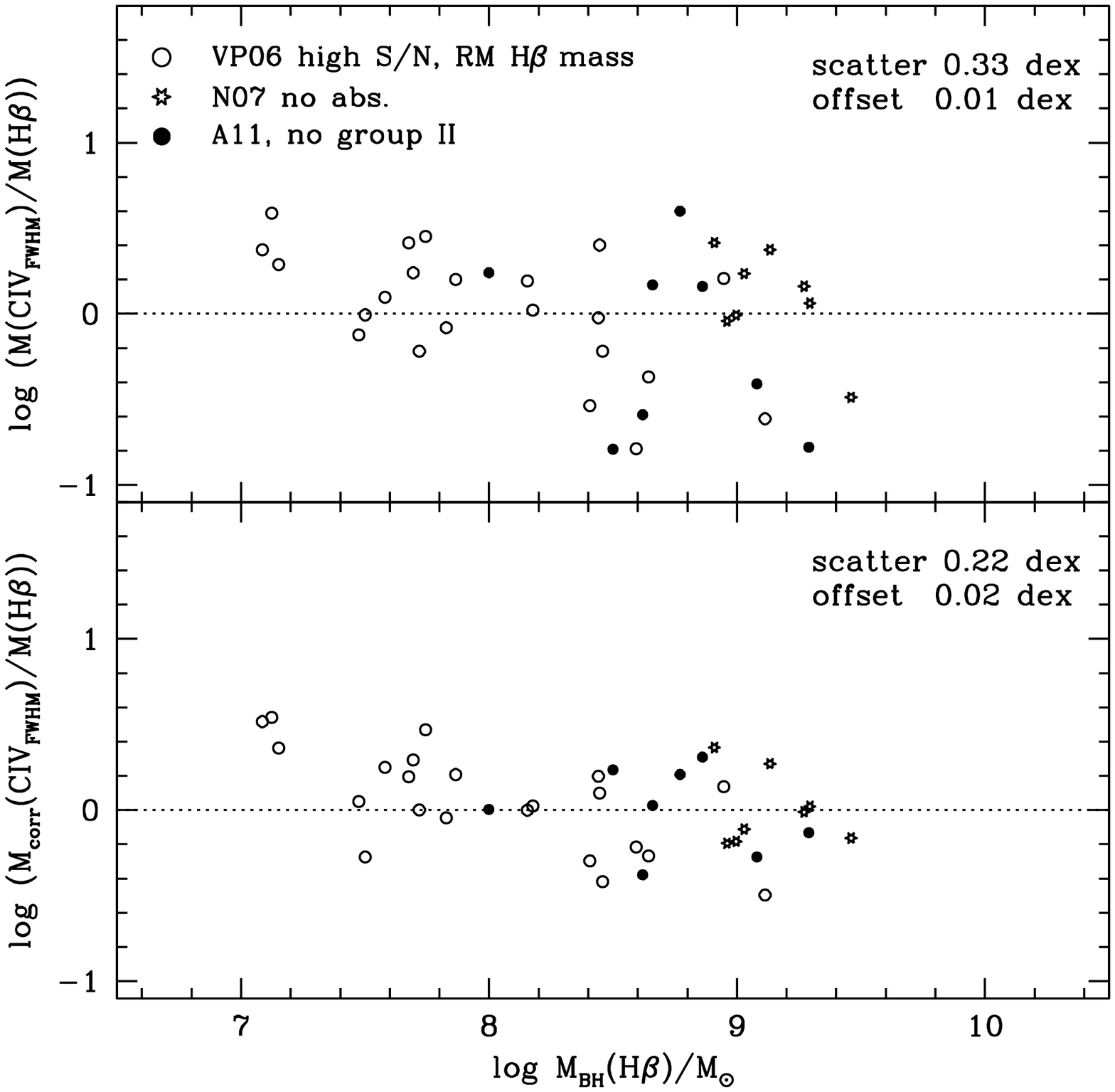}

\caption{Same as Figure \ref{Fig:origAcorrmassresid} except the top
  panel shows the subset of data kept from each sample after removing
  poor quality data, removing the A11 ``group II'' objects, and using
  the RM \Hbeta\ masses instead of SE masses for the VP06 sample.  The
  bottom panel shows this same ``high-quality'' sample after applying
  \civ\ mass corrections similar to Equation (1), but determined for
  each individual sample.  Mean sample offsets are listed but not marked
  because they are smaller than the typical measurement uncertainty on
  the mass residuals, $\sim 0.1-0.2$ dex for the VP06 sample.}

\label{Fig:bestorigAcorrmassresid}
\end{figure}

Finally, fully removing the data analysis inhomogeneity is not possible.
However, we fit and defined a shape-based \civ\ mass correction
independently for each sample, since the sample-to-sample
inhomogeneities in the spectral fitting and line width measurement
techniques have likely led to a sub-optimal \civ\ shape correlation fit
based on the combined sample.  We list the individual sample fit
parameters and uncertainties in Table \ref{Tab_Fits}. Figure
\ref{Fig:bestorigAcorrmassresid} shows the results of these
modifications to our analysis.  The top panel again shows the ratio of
the \civ-to-\Hbeta\ masses of this ``high-quality'' sample before
correcting \civ\ masses for the shape bias.  Here the scatter around the
1:1 relation is now 0.33 dex, and the systematic offset is small,
showing that simply removing poor quality data and host galaxy starlight
alone can reduce the scatter.  The bottom panel shows the residuals
after correcting the \civ\ masses using the individual shape corrections
determined for each homogeneous sample.  Here the scatter is only 0.22
dex, demonstrating the \civ\ and \Hbeta\ masses to be in very good
agreement.  The remaining scatter is less than the previously estimated
intrinsic systematic uncertainties of $\sim 0.3$ dex inherent in SE
masses (VP06) and is comparable to the uncertainty in the calibration of
the $R-L$ relationship (\citealp{Bentz09rl}, though see also
\citealp{Peterson10} for evidence that the intrinsic $R-L$ scatter is
somewhat lower).

\subsection{The BH Mass -- Luminosity Color Correlation}
\label{S:colorcorrelation}

In their comparison of \civ\ and \Hbeta-based masses in a sample of
lensed quasars, A11 discovered a strong correlation between the ratio of
\civ-to-\Hbeta\ masses and the ratio of UV-to-optical luminosity.  Such
a correlation is naively expected, since the ratio of these masses is a
function of the ratio of UV-to-optical luminosity.  However, the slope
of the correlation measured by A11 was in excess of that expected, if
the only source of the correlation was the dependence of $M_{\rm BH}$ on
$L$ ($M_{\rm BH} \propto L^{1/2}$).  Interestingly, if we combine the
blueshift--equivalent width relation of \citet{Richards11} with the
blueshift--color relation of \citet{Gallagher05}, we expect a
shape--color correlation in the sense that high (low) $S$ \civ\ profiles
are seen in bluer (redder) AGNs.  This means that our shape correction
operates in the same sense as the A11 color correction.

\begin{figure}
\figurenum{8}
\epsscale{1.0}
\plotone{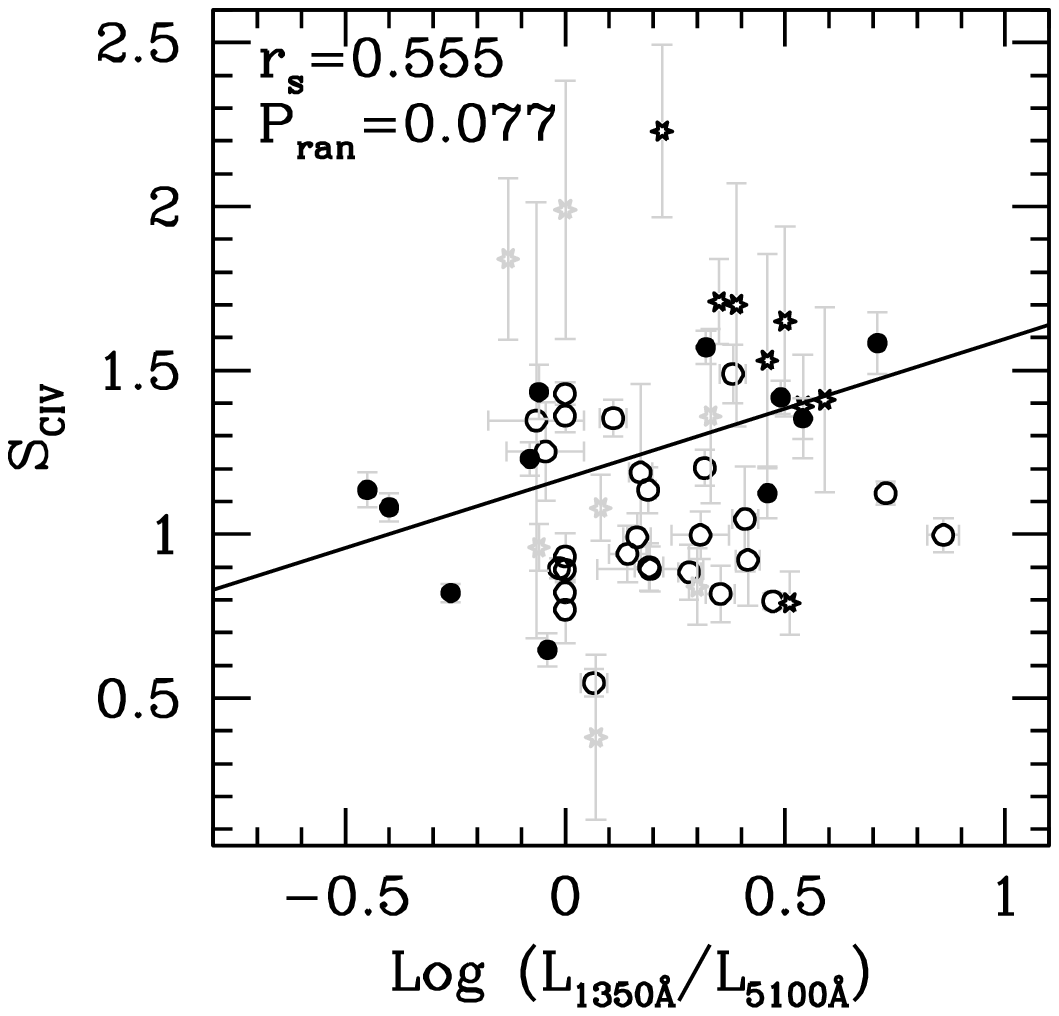}

\caption{Comparison between \civ\ line shape, $S={\rm FWHM}/\sigma_l$, and
  the ratio of the UV-to-optical luminosities,
  $L_{1350}/L_{5100}=\lambda L_\lambda\, (1350\aam )/\lambda L_\lambda\,
  (5100\aam )$. Symbol types are the same as in Figure
  \ref{Fig:massresidVshape}. The solid line is the best fit to the A11
  sample alone.  The statistics in the top left corner are for the A11
  sample only; the combined sample has $r_{\rm s}=0.231$ and $P_{\rm
    ran}=0.152$ (see Table \ref{Tab_Fits2}).}

\label{Fig:clrshape}
\end{figure}

Figure \ref{Fig:clrshape} shows this shape--color correlation.  To be
consistent, we used only SE, non-host-corrected $\lambda L_{5100}$ for
all objects, which necessarily decreased the VP06 sample by six objects.
We compare the statistics from this shape--color correlation with the
A11 mass--color correlation, but we first subtracted the expected
dependence of the A11 correlated quantities on the ratio of the
luminosities by removing the the luminosity term, $(\lambda
L_{1350})^{0.53}/(\lambda L_{5100})^{0.52}$, from the A11 mass
residuals.  This leaves only the excess correlation observed by A11, and
is a more consistent comparison with the shape--color correlation, in
which there is no redundant dependence on luminosity in both coordinate
axes.  The Spearman statistics listed in Table \ref{Tab_Fits2} show that
both correlations are weak, but the correlation of $S$ with color is the
stronger of the two.  Since A11 did not use this combined sample to
discover or define this correlation, we also make a comparison using the
A11 sample alone.  The fit to the A11 sample is denoted by the solid
line in Figure \ref{Fig:clrshape}, and the Spearman statistics are
listed the top corner as well as in Table \ref{Tab_Fits2}.  The
statistics show that these correlations are still not particularly
strong, but are more significant than for the combined sample (as also
noted by A11). The correlation of $S$ with color is, again, the more
statistically significant of the two.

\begin{deluxetable}{llccc}
\tablecolumns{4}
\tablewidth{0pt}
\tablecaption{Spearman Statistics for Correlations with UV-to-Optical Luminosity Ratio}
\tablehead{
\colhead{Correlated}& \multicolumn{4}{c}{}\\
\colhead{Quantity}&\colhead{Data Set}&\colhead{$r_{\rm s}$}&
\colhead{$P_{\rm ran}$} & \colhead{$N$}}

\startdata
$S$   &   Combined  &  0.231  &  0.152  &  40\\
$S$   &   A11       &  0.555  &  0.077  &  11\\
$\log (M_{\rm CIV}/M_{{\rm H}\beta}) -$ &&&& \\
\phn $\log (L_{\rm UV}^{0.53}/L_{\rm opt}^{0.52})$ & Combined & 0.207 & 0.201 & 40\\
$\log (M_{\rm CIV}/M_{{\rm H}\beta}) -$ &&&& \\
\phn $\log (L_{\rm UV}^{0.53}/L_{\rm opt}^{0.52})$ & A11 & 0.527 & 0.096 & 11\\
\enddata

\label{Tab_Fits2}
\end{deluxetable}

We also measured the mass--color correlation slope of the A11 Group I
objects after correcting these FWHM-based \civ\ masses with Equation (1)
above.  A11 originally measured this slope to be $0.89 \pm 0.25$.  We
measure a new slope of $0.71 \pm 0.17$, which is marginally consistent
with the expected slope of $\sim$0.53 from the mass scaling relationship
dependence on $L$.  This suggests that the shape bias in the \civ\
masses is a likely driver for the A11 mass--color correlation and can at
least partially explain the steep slope observed by A11.  Additional
work is underway using a larger sample to either identify or rule out
additional effects, e.g., non-universal SEDs, that could also be
contributing to the excess slope observed by A11.

\subsection{Origin of the Non-variable C\,{\scriptsize IV} Component}

Our analyses in \S\S \ref{S_RMmnrms} and \ref{S:shapebias} have shown
(1) that the SE \civ\ profile is a composite of both a variable
(reverberating) component and non-variable component, and (2) that
object-to-object differences in the characteristics of the non-variable
component are a primary cause for the large scatter in FWHM-based SE
\civ\ BH masses as compared to \Hbeta\ masses.  Here we discuss possible
origins for these differences.

\subsubsection{The ``Traditional'' Narrow Line Region}
\label{S_nonBLR}

\begin{figure}
\figurenum{9}
\epsscale{1.0}
\plotone{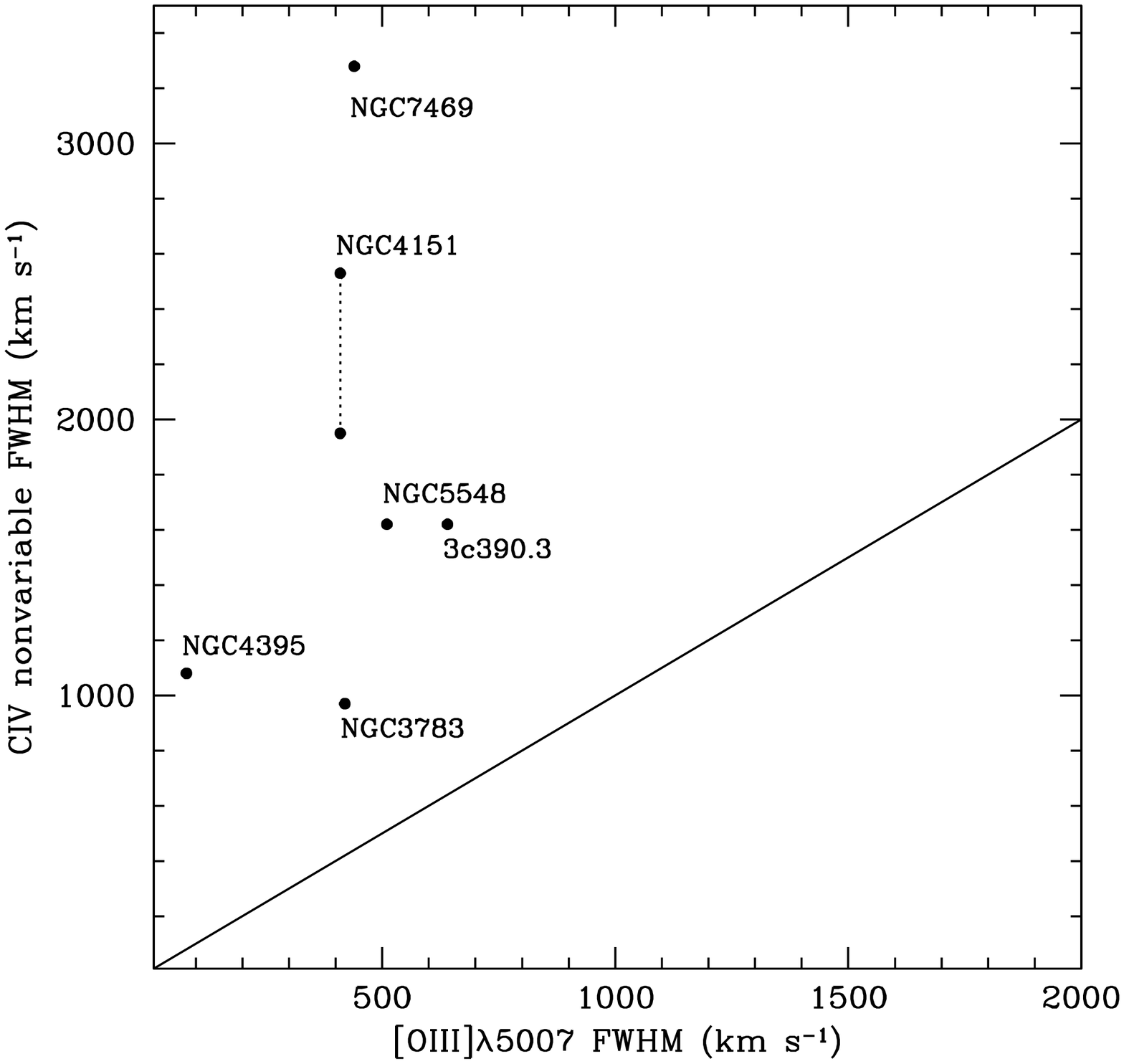}

\caption{Comparison between the FWHM of the \civ\ non-variable component
  and the FWHM of the [\oiii]\, $\lambda$5007 line for the RM sample.
  Individual objects are labeled, and the measurements reflect all
  objects shown in Figure \ref{Fig:civmnrms} except Fairall 9, whose rms
  spectrum was too noisy to accurately isolate the non-variable \civ\
  component.  The solid line is the 1:1 relation in FWHM, and the two
  \civ\ non-variable FWHM measurements for NGC\, 4151 connected by the
  dotted line represent the difference in width if the narrow absorption
  near 1545\AA\ is linearly interpolated across before measuring the
  FWHM (upper value) or not (lower value).}

\label{Fig:nonvarCIVvsOIIIwidths}
\end{figure}

The first, obvious, possibility is to associate the low velocity
component with the ``traditional'', low-density NLR, i.e., the region
responsible for emitting narrow forbidden lines, such as \ob.  A NLR
component is observed in the \Hbeta\ emission line and has sometimes
been assumed to be a component of the \civ\ profile as well
\citep[e.g.,][]{Baskin&Laor05, Sulentic07, Greene10, Shen12}.  However,
we argue that this cannot be the origin of the non-variable component of
\civ\ for the following reasons:

\begin{enumerate}
\item While the C+3 ion is photo-ionized, the \CIV\ emission line
  transition is excited collisionally.  Photoionization models suggest
  that \CIV\ emission from the low-density NLR region must therefore be
  weak \citep{Ferland&Osterbrock86}.  In contrast, the non-variable
  component seen in the mean \civ\ profiles of Figure \ref{Fig:civmnrms}
  is often quite strong.  Admittedly, \civ\ emission is still seen in
  Seyfert 2 spectra \citep[e.g.,][]{Collins05, Zheng08}, also suggesting
  a possible origin in the NLR, but studies have additionally shown that
  NLR emission line strengths and ratios are different between Seyfert
  1's and Seyfert 2's \citep{Zhang08} and may be dependent on
  inclination \citep{Fine11}.

\item Not all \civ\ SE profiles show an obvious narrow, core component.
  The presence and strength of this component is also anticorrelated
  with blueshift and the \civ\ equivalent width \citep{Wills93b,
    Richards02, Leighly04a, Richards11}.  It is much more probable that
  this reflects differences in the BLR kinematics or geometry between
  objects rather than correlations between NLR and BLR kinematics.

\item A large percentage of \civ\ SE profile peaks are blueshifted
  \citep{Richards11}, some by thousands of km s$^{-1}$.  While NLR
  emission can be blueshifted from the systemic redshift, it would {\it
    not} be expected to show these large of blueshifts.  If present, NLR
  emission should appear at or at least near the systemic redshift,
  leaving it relatively redshifted with respect to the highly
  blueshifted \civ\ profile \citep{Wilkes84}.  One would then expect
  potential double-peaked profiles in these objects, rather than this
  component simply being absent.  Such a feature is not observed here,
  in large samples, or even in composite AGN spectra \citep[see,
  e.g.,][]{Richards11}.

\item We subtracted the scaled rms spectrum from the mean spectrum of
  all objects shown in Figure \ref{Fig:civmnrms} (except Fairall 9
  because of the poor quality rms spectrum) and measured the FWHM of the
  residual non-variable components using the same methods as in Section
  \ref{S_RMmnrms}.  We also measured the FWHM of the
  \oiii\,$\lambda$5007 line from this sample, subtracting a linear,
  local continuum, from available optical data.  Figure
  \ref{Fig:nonvarCIVvsOIIIwidths} shows that the velocity widths of the
  non-variable narrow cores are larger (often significantly so) than
  those measured in the unblended NLR \oiii\,$\lambda$5007 line.
  Combining these relatively broad core widths with the core strength
  and additional blue excess emission in most of this sample, it is
  unlikely that this component is, at least solely, from the
  traditional, low-density, extended NLR emission.

\end{enumerate}

\subsubsection{A BLR Disk-wind}
\label{S_DWBLR}

The presence of a BLR wind has been used to explain a wide variety of
observed AGN phenomena.  This includes, for instance, the often
blueshifted \civ\ profile \citep[see][and references
therein]{Richards11}, absorption line features
\citep[e.g.,][]{Hamann11}, and AGN outflows and feedback
\citep[e.g.,][]{Reeves09}.  \citet{Richards11} argue that the sources
with the largest blueshifted \civ\ profiles are the ``wind-dominated''
sources, where this wind component causes a blueshifted peak and line
asymmetries \citep[cf.][]{Leighly04a} in the SE \civ\ profile.  These
observations are consistent with the non-variable blue excess emission
we observe when comparing the RM mean and rms spectra.  However, we also
observe a strong low-velocity, non-variable core in some objects.  So in
terms of the SE BH mass estimates {\it alone}, if this total
non-variable component (core $+$ blue excess) is a disk-wind, it is the
peaky, low $S$-value objects with the strongest SE core components that
would have the largest wind ``contamination'' or bias in their FWHM and
therefore FWHM-based SE mass measurements.

The BLR disk-wind has been suggested to be launched from the inner, {\it
  high-velocity}, high-ionization regions of the BLR or outer accretion
disk \citep[cf.][]{Murray&Chiang97, Elvis00, Elvis12}, so the additional
non-variable, {\it low-velocity} core component observed here does not
readily fit into this model unless there is an orientation dependence of
the wind component relative to the line of sight (LOS).  The RM mean and
rms spectra in Figure \ref{Fig:civmnrms} as well as \civ\ profiles in
general show that the strength of the blueward asymmetric emission
appears to be roughly anti-correlated with the strength of the
low-velocity core (e.g., the blueshift--equivalent width relation
discussed by \citealt{Richards11}).  Therefore, we argue that if this
non-variable component does originate in an outflowing BLR wind, there
{\it should} be an orientation dependence, where the change in \civ\
profile from `peaky' to `boxy', or low $S$ to high $S$, is due to
varying levels of LOS wind contamination to the broad emission line
profile as a function of orientation.  Furthermore, because this
component does not reverberate, such a wind would necessarily be
optically thin to the ionizing continuum \citep[see, e.g.,][who discuss
the possibility that low optical depth can explain the non-variable
wings of the \Hbeta\ profile]{Korista&Goad04}.

For a disk$+$wind BLR \citep[cf.\ Figure 1 of][]{Murray&Chiang97} to
support our observations of the variable$+$non-variable \civ\ profile
and put these observations in context to other \civ\ observations, an
optically-thin wind that is more polar than equatorial is required.  In
this case \CIV\ can still be emitted in the wind but does not
reverberate.  We can then make a simple generalization for a composite
``disk'' (variable component) plus ``wind'' (non-variable component)
\civ\ profile \citep[see also][]{Wang11}.  We would then expect the
following orientation-dependence of the profile of each component:
\begin{itemize}

\item {\bf The disk:} As the inclination decreases (goes from edge-on to
  face-on), the disk component experiences a general narrowing in
  velocity width due to the change in inclination.  There is no other a
  priori requirement for the shape of the line; photoionization and BLR
  geometry model-fits to the Balmer line profiles have shown that a wide
  variety of both single- and double-peaked profiles can exist and still
  arise from a virialized disk-like distribution of gas with varying
  optical depths \citep[see review by][and references
  therein]{Eracleous09}.

\item {\bf The wind:} At high (closer to edge-on) inclinations, the wind
  component is narrow and centered near the \civ\ systemic velocity, as
  the direction of outflow is largely perpendicular to the LOS.  As the
  inclination decreases, more of the wind is observed to be outflowing
  along the LOS.  Consequently, the wind emission is distributed across
  a larger, preferentially blue-shifted, range of velocities, so the
  peak also blueshifts and decreases in flux.

\end{itemize}

\subsubsection{The Intermediate Line Region}

Another previously suggested explanation for the peaky SE core component
in some \civ\ lines is that this emission arises in an ``intermediate
line region'' \citep[ILR;][]{Wills93b, Brotherton94}.  Such a region is
explained as a higher-velocity, higher-density inner (i.e., closer to
the BLR) extension of the NLR.  \citet{Wills93b} alternately suggest a
different, simple model in which the core \civ\ emission could be coming
from a bipolar outflow, but this may not be mutually exclusive to the
ILR/inner NLR.  More recent spatially-resolved studies of the kinematics
and physical conditions in the inner NLRs of local Seyferts
\citep[e.g.,][]{Das05, Kraemer09, Crenshaw10, Fischer10, Fischer11} show
a link between possible ILR \civ\ emission and biconical outflows
\citep[cf.][]{Crenshaw&Kraemer07}.  While we mentioned above that
densities in the ``typical'', i.e., extended, NLR are not high enough to
result in significant emission from high-ionization species, spatially
resolved STIS spectra have co-located high-ionization-line emission with
knots of [\oiii] $\lambda5007$ emission in the inner NLR
\citep[e.g.][]{Collins05}.  Such emission must be enhanced by
collisional processes \citep{Kriss92}, such as shock heating or
microturbulance, mechanisms likely to be present if this emission is
arising in an outflow \citep{Kraemer07}.  \citet{Nelson00} and
\citet{Kraemer00} show that with {\it HST} observations of nearby
Seyferts, spatially-resolved \CIV\ emission does arise outside the
unresolved nucleus, i.e., from the ILR/inner NLR.  However, the \civ\
emission flux drops significantly faster than \ob\ emission as a
function of radius.  The spatially resolved inner knots of [\oiii]
$\lambda5007$ emission also show relatively high, $\sim$~1000 km
s$^{-1}$, and even blueward asymmetric velocities \citep{Nelson00,
  Das05, Fischer11}, both reminiscent of the non-variable \civ\
component we see here.  Unfortunately, and important for potentially
characterizing the non-variable \civ\ profile, because the [\oiii]
emission extends to much larger radii and lower velocities, the line
profile of the ``integrated'' [\oiii] emission in typical ground-based
spectra of both near and distant objects is not a good template match
for the observed non-variable \civ\ core emission.

In this type of scenario, the non-variable component of the \civ\ line,
i.e., both the peaky, low-velocity \civ\ core emission and blue excess,
could be attributed to the ILR emission and part of a biconical outflow.
LOS emission-line profiles from such a distribution of gas would
therefore exhibit velocity widths larger than the typical, extended NLR
region that is probed by integrated forbidden-line emission profiles,
but narrower than typical BLR emission.  Additionally, if the gas is
outflowing, blueshifts are to be expected, and potentially more
prominent for the higher gas velocities, since this emission likely
arises at closer radii and may therefore by more highly collimated by
the bicone.  These are all expectations consistent with the observations
of the \civ\ emission line profile seen here and in previous studies.
Furthermore, our orientation-dependent model of the non-variable \civ\
emission introduced above fits equally well with an origin in a
biconical ILR outflow as in the BLR disk-wind, since the LOS component
from the inner region of such an ILR outflow should change as a function
of orientation.  Such a distribution of gas would also not be expected
to reverberate.  Because the spatial scales are large relative to the
time scale of variability, any reverberation signal would be spatially
damped.  Therefore, the non-variable nature of such an emission
component naturally holds for this origin, at least on reverberation
time scales.

\subsection{Evidence for Orientation-Dependence}

Both origins for the non-variable \civ\ component that we suggest above
require an orientation dependence to the outflow component of \civ\
emission.  We first looked for indications of an orientation dependence
by comparing our orientation expectations with the \civ\ profile trends
observed in the RM sample.  To do this, however, we must isolate the
non-variable component profile to search for signs of orientation
dependence.  This was done through a spectral decomposition of the mean
spectrum \civ\ profiles shown in Figure \ref{Fig:civmnrms}.  We assume
that the mean profile is a superposition of the BLR `disk' (variable
component) and outflow (non-variable component).  The rms profile shows
variable emission only, so we generalize this emission to arise solely
in the reverberating BLR `disk'.  We decompose the mean profiles into
variable and non-variable components by first fitting the scaled rms
profile (solid gray curves in Fig.\ \ref{Fig:civmnrms}) with a sixth
order Gauss-Hermite polynomial as a means to derive a smooth profile for
the disk component.  We subtract this component from the mean spectrum,
and attribute the residual non-variable flux to the outflow component.
We fit the residual emission again with a GH polynomial after
interpolating over any narrow absorption features.

\begin{figure}
\figurenum{10}
\epsscale{0.9}
\plotone{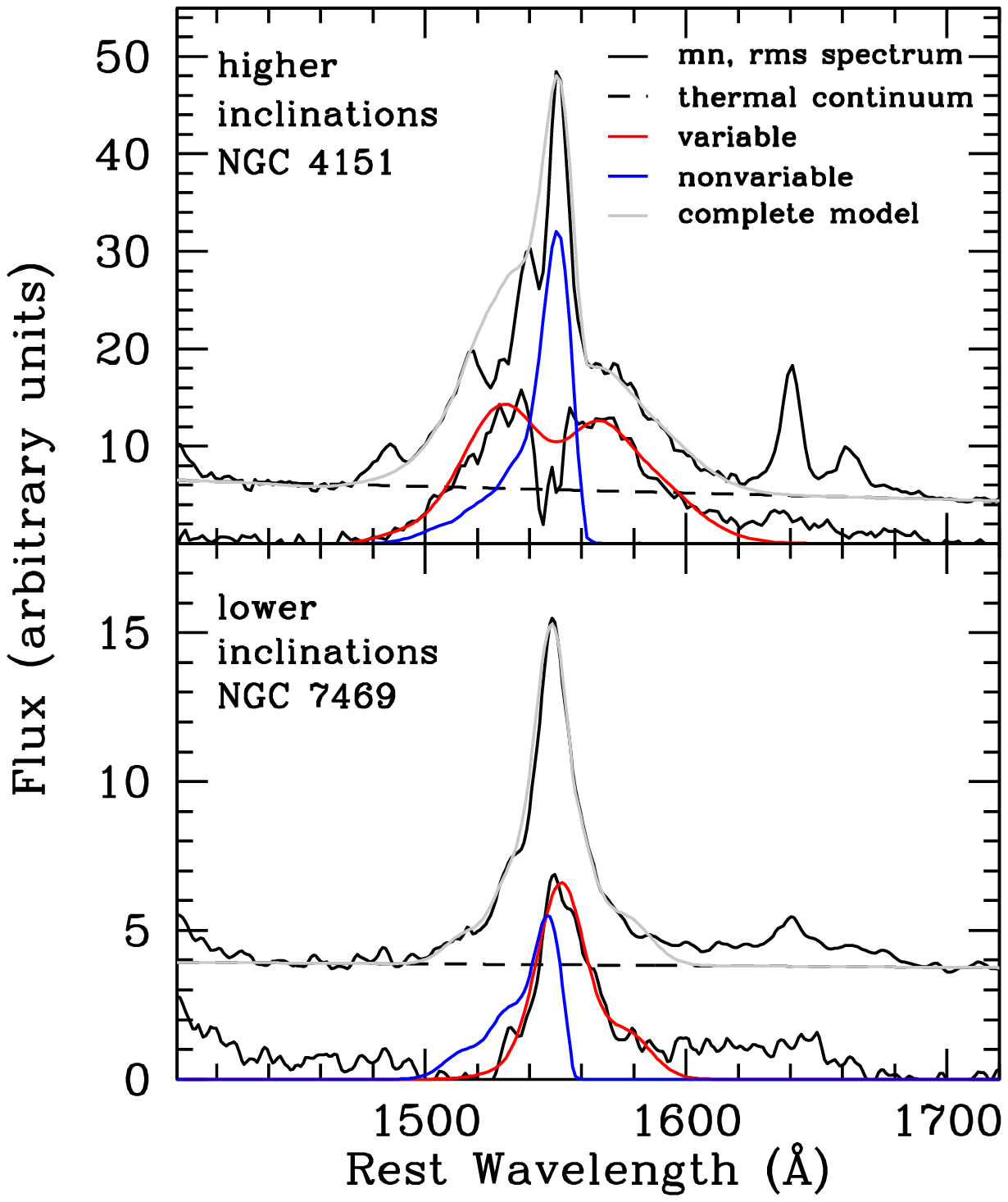}

\caption{Predicted emission line profiles expected if observing an
  object with a disk-wind BLR at higher, more edge-on inclinations,
  ({\it top}) and lower inclinations ({\it bottom}).  For each panel,
  the higher flux black curves show the original mean spectrum (also
  shown in Fig.\ \ref{Fig:civmnrms}) used to represent fiducial \civ\
  profiles for each orientation.  The red curve represents the variable
  model component fit to the scaled rms spectrum (also in black), the
  blue curve represents the non-variable model component, and the gray
  curve is the composite model including both components.  Finally, the
  dashed curve is the AGN continuum that was subtracted before fitting
  each emission-line component.}

\label{Fig:diskwindprofile}
\end{figure}

We highlight this decomposition for the \civ\ profiles of NGC\, 4151,
whose mean spectrum has a relatively low $S$-value (0.61) with a strong
non-variable core, and NGC\, 7469, which has higher $S$ (0.92) with a
notably weaker non-variable core but with additional, non-variable blue
excess.  The results of the decomposition for these two sources are
shown in Figure \ref{Fig:diskwindprofile}, where the original mean and
rms spectrum and individual and composite (disk plus outflow) profile
fits are shown for NGC\, 4151 and NGC\, 7469 in the top and bottom
panels, respectively.

From this exercise, we observe for NGC\, 4151 that (1) the rms profile
(our disk component) is broad and double-peaked, possibly demonstrating
the emission expected from an inclined disk \citep{Eracleous&Halpern94},
and (2) the relatively narrow velocity distribution of non-variable
residuals would also be expected from an outflow directed largely
perpendicular to the LOS.  While the model-fit to the rms spectrum
over-predicts the flux in the core, this acts only to somewhat
underestimate the core strength but does not significantly change the
profile of the residual non-variable component.  As it would not be
possible to view this broad-line AGN perfectly edge-on and still see a
broad-line AGN (in the unified model), the extended and blue-shifted
wing in the non-variable component profile is also consistent with the
expectations for this component arising in an outflow.  For NGC\, 7469
we see (1) a less peaky and more blueshifted outflow component and (2) a
narrower disk component.  These are both expected characteristics of a
more face-on orientation.  The combined fit conspires to produce a
smooth but slightly asymmetric profile, characteristic of many SE \civ\
profiles.

The orientation-dependent model we describe here is naively simple.
Interestingly, however, our crude decomposition using two test cases
from the RM sample places no a priori expectations for the profiles of
the individual variable and non-variable components.  Yet, the results
of attributing these components to a BLR disk$+$outflow composite
profile are strikingly similar to what is expected if we are viewing
NGC\, 4151 at a higher inclination (more edge on) and NGC\, 7469 at a
relatively lower inclination (more face on).

There are also numerous discussions in the literature of orientation
dependencies to observed AGN properties.  For example,
\citet{Richards02} suggested the possibility of an orientation effect in
the \civ\ blueshifts, and thus wind component, though interestingly in
the opposite sense that we argue here.  \citet{Richards02} advocated
that large blue-shifted profiles were viewed from more edge-on
orientations, which follows the expectations of the largely equatorial
wind in the \citet{Murray&Chiang97} model.  However, \citet{Richards11}
argue against such a dependence based on radio morphology studies.
Other radio morphology studies also show no or only weak
orientation-dependence of \civ\ widths \citep{Jarvis&McLure06, Fine11}.
However, these studies are based primarily on using the FWHM, and we
have shown here that \civ\ FWHM measurements are biased by the
non-variable component.  In fact, \civ\ profiles resulting from an
orientation-dependence of the non-variable component in the sense we
propose above (creating preferentially peaky profiles for high
inclination and boxy profiles for low inclination) would act to mask the
expected changes in the FWHM of the BLR `disk' component that these
studies hope to probe:  The non-variable component contamination will
lead to an artificially large measurement of the FWHM for low
inclinations and an artificially small measurement of the FWHM for high
inclinations.

Another place to search for possible orientation-dependent quasar
phenomena is in the characteristics of broad absorption line quasars
(BALQSOs), which have been inferred to be viewed preferentially more
edge-on \citep[e.g.,][]{Murray95, Elvis00}.  This is partly because
their continua are relatively more reddened than non-BALQSOs
\citep[][and references therein]{Reichard03}.  However, other
observations and models have shown that (1) polar outflows are also
capable of producing BALs \citep{Brotherton06, Zhou06, Ghosh07,
  Borguet10}, (2) individual fits to BALs cannot constrain the geometry
and orientation of the BAL region \citep{Hamann93}, and (3) radio BAL
and nonBAL quasars are nearly indistinguishable across a wide range of
observed and physical properties, including spectral shape, spectral
index, and polarization properties \citep{Bruni12}.  This evidence
suggests that BALs can be observed in quasars over broad range of
orientations.

Furthermore, discussing the results of some previous studies in the
context to this simple orientation dependence of the \civ\ SE profile
leads to some novel interpretations of the observations (though we do
not claim to be able to explain every aspect of these observed phenomena
with {\it only} inclination):
\begin{itemize}

\item The Baldwin Effect \citep{Baldwin77}. We argue that the lowest
  equivalent width objects are those that tend to have high $S$ and are
  the most face-on.  The continuum photons escaping the central source
  should therefore be the least obscured by other nuclear material.
  This will lead to a larger observed luminosity, on average, compared
  to higher inclination objects, which we argue are observed to have
  high-equivalent width, low $S$ profiles due to wind contamination.

\item NLR Component Removal \citep[e.g.][]{Baskin&Laor05, Sulentic07}.
  Some studies argue the narrow \civ\ core originates in the NLR.  Yet,
  \citet{Sulentic07} have pointed out that the narrow-line components in
  \civ\ are often broader and stronger than typical, ``\oiii\,
  $\lambda$5007-like'', NLR emission.  Though \citet{Sulentic07} found
  better correlations between \civ\ and \Hbeta\ after removing a narrow
  core component, they were still left with a large scatter between
  their \civ\ and \Hbeta\ masses.  In the context of our model and other
  arguments in Section \ref{S_nonBLR}, the large scatter remaining in
  the \citet{Sulentic07} results is most probably due to the uncertainty
  in fitting and subtracting their ``narrow'' component.  Since the
  orientation-dependent non-variable component cannot simply be
  described as a typical NLR emission line, a traditional spectral
  decomposition does not work for \civ.

\item ``Extincted'' red wing of \civ\ \citep[see e.g., composite spectra
  of][]{Richards02, Richards11}.  Past physical explanations for this
  phenomenon \citep[cf.][]{Gaskell82} include a combination of radial
  gas motions and obscuration, though Gaskell admits that it is
  questionable whether the dust grains required for such obscuration
  could survive in the BLR.  Our orientation-dependent disk-outflow
  model explains this phenomenon simply: as the inclination decreases,
  the observed range of disk velocities narrows.  Conversely, the
  observed velocities of the outflow component are increasing as it
  becomes directed more along the LOS, effectively filling in the
  high-velocity portion of the profile previously occupied by the disk
  emission.  However, this process is preferentially blueward asymmetric
  because the receding component of the outflow is not visible through
  the disk.  This leaves the red side of the profile devoid of emission
  from the highest velocity gas.

\item Correlation of blueshift with luminosity \citep{Richards11}. This
  is expected if these are the population of objects seen more face-on.
  Furthermore, \citet{Gallagher05} demonstrated with a color-blueshift
  distribution that large blueshift AGNs are less likely to have red
  continua, regardless of whether intrinsic or dust-reddened.  This
  would similarly be true if we are seeing these objects more face-on.

\item Intrinsic colors of BALQSOs \citep{Reichard03}.  Despite the
  observed trend of BALQSOs being more reddened than nonBAL quasars,
  \citet{Reichard03} found that after correcting for SMC-like dust
  reddening, BALQSOs appear to be drawn from the same parent population
  as nonBALs \citep[see also][]{Bruni12}.  Upon further detailed
  inspection of BALQSO properties as a function of `intrinsic' colors,
  \citet{Reichard03} also discovered the same nonBAL quasar \civ\
  emission line trends of larger (smaller) equivalent width profiles for
  `intrinsically' red (blue) HiBALs (their Figure 9).  This suggests to
  us that the obscuration typical of BALQSOs may not necessarily be an
  orientation dependent effect (e.g., from a torus, which would indicate
  a preferential edge-on orientation).  Instead, if the \civ\ line shape
  trend {\it is} an orientation effect as we suggest, the obscuring
  regions responsible for producing BALs in quasars may in fact be
  orientation independent. The range of observed \civ\ profiles in
  BALQSOs is then additional evidence that BALs can be observed over a
  range in orientations.

\end{itemize}

Realistically, there are many geometrical, kinematical, and physical
properties of the BLR that could cause varying contributions from the
LOS component of a BLR disk-wind, and most likely, more than one effect
is present in varying degrees for different samples of AGNs.  For
example, \citet{Murray&Chiang97} find that stronger core emission from
an equatorial wind component can occur without requiring an
inclination-dependence; instead, it is the result of simply extending
the outer radius of the wind-emitting region.  We focus only on
inclination here because it is arguably the simplest to conceptualize,
still suffices to explain many observed trends, and fits the
observations whether the non-variable emission originates in an outflow
from the BLR or ILR.  Ideally, we would like to use direct orientation
measurements to corroborate our interpretation that the \civ\ profile
changes are a result of an orientation-dependence of the non-variable
emission.  Unfortunately, such orientation measurements are scarce.

The most promising methods use radio morphology
\citep[e.g.][]{Brotherton96, Vestergaard00, Jarvis&McLure06, Fine11} or
NLR kinematics \citep[see, e.g.,][]{Fischer10, Fischer11}.  As mentioned
above, past radio studies may be biased by typically only using the
FWHM, and these studies should be revisited \citep[though see][who find
the strongest \civ\ orientation dependence using the full width at
$20\%-30\%$ peak intensity, which would be less susceptible to
contamination from the non-variable core]{Vestergaard00}.  NGC\, 3783
and NGC\, 4151, two objects in our RM sample, have been targeted for NLR
kinematical modeling to determine orientation.  The best-fit models
indicate an inclination of only $15^{\circ}$ away from the LOS for NGC\,
3783 (T.\ Fischer, priv.\ comm.; Fischer et al.\ in prep) and
$45^{\circ}$ for NGC\, 4151, near the maximum expected for a Type 1
spectrum \citep{Das05}.  Putting the relative strength of the
non-variable core observed in each of these objects in Figure
\ref{Fig:civmnrms} in context to these independent orientation estimates
seems to support our expectations that we are observing NGC\, 3783 and
NGC\, 4151 at relatively low and high inclination, respectively.

\section {Summary and Conclusion}
\label{S_conclusion}

We have investigated sources for the large observed scatter between
\civ- and \Hbeta-based single-epoch AGN black hole masses.  Our primary
line of investigation was to return to the assumptions on which SE BH
masses are based by comparing the \civ\ profiles in the mean and rms
spectra from reverberation mapping experiments.  These include the
requirement that the velocity width of the mean, or SE profile be a
reasonable representation of the rms profile width --- the velocity
profile corresponding to emission from reverberating gas.  We found that
the SE or mean \civ\ emission-line profiles often have a non-variable
low-velocity core component and/or blue excess that does not appear in
the rms profiles and therefore does not reverberate.  The presence of
these non-variable components in the SE \civ\ profiles are biasing
\civ-based BH mass determinations.

We developed an empirical correction for this bias, applicable to
FWHM-based \civ\ masses by using the observed characteristics of the
\civ\ profile, which we parameterize through the ratio, $S={\rm
  FWHM}/\sigma_l$, of the \civ\ FWHM to $\sigma_l$.  After applying this
empirical ``shape'' correction to the FWHM-based \civ\ SE BH masses in a
sample of $\sim50$ low- and high-redshift AGNs with both \civ\ and
\Hbeta\ mass estimates, we found that the scatter in the \civ-to-\Hbeta\
mass residuals is reduced from 0.39 dex to 0.26 dex for this somewhat
heterogeneous sample.  

We then investigated the remaining scatter in the \civ-to-\Hbeta\ mass
ratio by looking at other possible sources.  We concluded that data
inhomogeneity, poor data quality, and independent biases in the \Hbeta\
masses due to NLR emission and host galaxy starlight also contribute to
the observed scatter.  If we minimize these systematic problems, we
found that we could further reduce the scatter in the \civ-to-\Hbeta\
mass residuals to 0.22 dex.  The significant additional decrease in this
scatter as a result of applying individual shape correction factors to
each smaller, more homogeneous sample suggests that a shape correction
calibration determined from a highly homogeneous sample of both \civ\
and \Hbeta-based SE masses may further reduce the mass residual scatter
compared to that seen here.  We are currently in the process of
collecting such a sample of high $S/N$, simultaneous observations of
both the \civ\ and \Hbeta\ emission line regions in a large sample of
AGNs using the Xshooter spectrograph on the VLT \citep[see e.g.,][for a
comparatively smaller sample of such data]{Ho12}.

Finally, we investigated whether the \civ\ shape trend could also be
responsible for the mass--color correlation described by A11.  We found
only a weak correlation between $S$ with this UV-to-optical color term.
However, this weak shape--color correlation had a similar slope but was
actually tighter than the A11 mass--color correlation after removing the
expected dependence of that correlation on $L$, implying that the
mass--color correlation is simply a manifestation of the \civ\ shape
bias.  We further supported this interpretation by demonstrating that
the steep slope observed by A11 in the mass--color correlation largely
disappears after applying the shape-correction defined in Equation (1)
to the A11 FWHM-based masses.

Our shape parameterization of the \civ\ line profile and correction to
the FWHM-based \civ\ masses was strictly empirical in construction, and
not altogether dissimilar from some previous studies
\citep[e.g.,][]{Baskin&Laor05}.  However, the additional results gleaned
from the reverberation database of mean and rms \civ\ spectra led to the
discovery that the observed range of SE \civ\ line shapes correlates
with the presence of and object-to-object differences in a non-variable
emission component.  The physical interpretation we suggest could
explain the observed differences in this non-variable component is via
orientation-dependent sight-lines of a (largely) polar outflow.  The SE
\civ\ emission-line profile predictions from such a model (peaky in more
edge-on sources, boxy and blue-shifted in more face-on sources) explains
the distribution of observed SE profiles as composites of both the
reverberating BLR emission and ``contamination'' from the
orientation-dependent non-variable outflow component.  This
contamination biases the velocity width measured in the SE spectrum
relative to the intrinsic velocity dispersion of the reverberating gas,
thus biasing SE mass estimates.

The radial origin of such an outflow remains an unanswered question,
however.  Profile and velocity width comparisons of this non-variable
component with the [\oiii] $\lambda$5007 emission line suggest it must
be arising from closer in than the extended NLR probed by {\it
  integrated} forbidden line emission.  However, it is difficult without
further observations and advances in modeling to conclude whether the
outflowing gas producing the non-variable \civ\ emission originates in
an inner BLR disk-wind, the intermediate line region, aka, high-velocity
inner extension of the traditional NLR, or even a combination of the
two.  At present, we show no preference as to what that origin may be.
Certainly an origin in the intermediate line region produces natural
mechanisms to explain the lack of variability of this component (spatial
damping) and the orientation-dependence (collimation in a biconical
outflow). Plus, spatially resolved biconical outflows have been shown to
exist on these scales.  On the other hand, such an outflow must be
launched from somewhere within the unresolved nucleus.  Models seem to
favor the inner (high-ionization) regions of the BLR or outer accretion
disk for the ultimate source of AGN outflows/winds, but the
orientation-dependence of such a wind is, at present, model-dependent.

Future observational studies could potentially constrain the physical
origin of this component by studying any potential time dependence of
the equivalent width and velocity profile of this component.  For
example, gross changes in the broad emission-line profile are observed
to exist over dynamical time scales for the size of the BLR \citep[see,
e.g.,][]{Sergeev07}.  So, while we have concluded that this outflow
component does not {\it reverberate}, if the origin of this component is
from an intermediate line region, variability may still be expected on
much longer time scales than a single reverberation campaign.  If
present, a measure of this variability time scale could put a constraint
on the location of this emitting region.  On the other hand, if the
emission is originating in an optically thin wind close to the central
source, changes in the amplitude of emission may be expected with
changes in the AGN luminosity state.  As the ionizing continuum
luminosity increases and the ionization front moves outward, we may
expect an increase the equivalent width of this emission because the
volume and temperature of the ionized gas will also increase.

Regardless of the physical origin, this non-variable component is likely
present in some strength and form in all SE \civ\ emission lines, and we
have shown it to be a large, but correctable, source of systematic bias
in \civ-based SE BH mass estimates.  While we suggest the possibility
that the profile of this non-variable component is orientation
dependent, more sophisticated models of the BLR and AGN central engines
\citep[cf.\ the models of][]{Chiang96, Murray&Chiang97, Proga00,
  Proga04, Ohsuga11, Proga12} taking into account not only orientation,
but also photoionization physics, kinematics, geometry, and even
radiation pressure are needed.  Such models will help us to better
understand the full complexity and origin of the \civ\ emission (both
reverberating {\it and} non-reverberating) as well as other observed AGN
properties.  Quasar evolution and/or SED and $L/L_{\rm edd}$ changes
with luminosity could also affect the relationship between the variable
and non-variable \civ\ emission components.  Such changes could be
independent of orientation and/or lead to different orientation
dependencies as a function of these parameters, if, for instance, these
physical properties effect the bicone opening angle or the driving
mechanism for AGN outflows.  If such models and/or additional
observational tests can corroborate inclination as the source or at
least a large contributor to the differences in the observed \civ\
profiles and for either all or even particular samples of AGNs, having
an inclination indicator as simple as the observed SE \civ\ profile
could be a very powerful tool for better understanding the physics of
AGNs and reducing systematics in both SE and RM masses.

As an interesting side note, if a source of this non-variable \civ\
emission is a BLR disk-wind, we may also consider the possibility that
the non-variable, high-velocity wings of \Hbeta\ that disappear in rms
spectra and have also been potentially attributed to emission from
optically-thin gas \citep[see discussion by][]{Korista&Goad04}, may also
be from this same BLR wind component.  \Hbeta\ is emitted from larger
BLR radii than \civ, so it may only be the very highest velocities of
\Hbeta-emitting gas that are affected by this wind component,
particularly as it has been theorized to be launched from the inner BLR.

In the end, however, the rms spectra derived from reverberation mapping
campaigns are the only guaranteed means by which to directly probe the
uncontaminated, variable profile most relevant for determining accurate
black hole masses and to decompose the non-variable component by which
we can study its origin.  Therefore, more reverberation mapping
experiments targeting objects across the full observed distribution of
SE \civ\ profiles as well as multiple epochs for individual objects are
needed (1) to expand the sample of \civ\ mean and rms spectra, (2) to
form a more robust \civ\ shape correction, and (3) to better understand
the extent and physical origin of this non-variable component and the
bias it imposes on \civ-based BH mass estimates.  Furthermore, these
programs must also probe larger AGN luminosities and higher redshifts so
that results can more directly be applied to the typical quasar
populations observed in surveys.  Currently, our only option is to
extend what has been learned for the relatively, low-luminosity, local
Seyfert sample presented here.  However, it is imperative for our
understanding of galaxy evolution to directly test what systematics, if
any, we are introducing by extending results from these local
investigations to AGNs at higher redshifts, luminosities, and even in
potentially different evolutionary states.

\acknowledgements 

I would like to graciously thank R.~J.~Assef, C.~S.~Kochanek,
S.~B.~Kraemer, B.~M.~Peterson, G.~T.~Richards, and M.~Vestergaard for
stimulating and insightful discussions, helpful comments, and
thought-provoking critiques that improved both the content and
presentation of this work.  I would also like to thank B.~M.~Peterson
and S.~Kaspi for providing \civ\ spectra of the reverberation mapping
sample presented, and T.~C.~Fischer and D.~M.~Crenshaw for the NGC\,
4395 optical spectrum.  Finally, I would like to thank the anonymous
referee who provided suggestions that further improved the clarity of
the discussion and conclusions.  Discussions during the workshop {\it
  Improving Black Hole Masses in Active Galaxies} that took place 2012
July in Copenhagen, Denmark, helped improve the content of this work.
The research leading to these results has received funding from the
People Programme (Marie Curie Actions) of the European Union's Seventh
Framework Programme FP7/2007-2013/ under REA grant agreement no.\
300553.  The Dark Cosmology Centre is funded by the Danish National
Research Foundation.





\end{document}